\documentclass[aps,twocolumn,superscriptaddress]{revtex4}
\usepackage{amsfonts}
\usepackage{amsmath}
\usepackage{amssymb}
\usepackage{graphicx}
\usepackage{xcolor}
\usepackage{float}
\usepackage{lipsum}

\begin{document}

\title{Non-Hermitian optical scattering in cold atoms via four-wave mixing}
\author{Xiao Liu}
\affiliation{School of Physics and Center for Quantum Sciences, Northeast Normal University, Changchun 130024, China}
\author{M. Artoni}
\email{maurizio.artoni@unibs.it}
\affiliation{Department of Engineering and Information Technology, Brescia University, 25133 Brescia, Italy}
\affiliation{European Laboratory for Nonlinear Spectroscopy $\& $ Istituto Nazionale di Ottica del CNR (CNR-INO), 50019 Sesto Fiorentino, Italy}
\author{G. C. La Rocca}
\email{giuseppe.larocca@sns.it}
\affiliation{NEST, Scuola Normale Superiore, 56126 Pisa, Italy}
\author{Jin-Hui Wu}
\email{jhwu@nenu.edu.cn}
\affiliation{School of Physics and Center for Quantum Sciences, Northeast Normal University, Changchun 130024, China}
\date{\today}

\begin{abstract}
Nonlinear effects could play a crucial role in addressing optical nonreciprocal behaviors in scattering media. Such behaviors are, however, typically observed within a single transmission channel and predominantly in media with fixed optical structures, which inherently restrict the tunability of a nonreciprocal response. We suggest to combine the (intrinsic) nonlinearities of a coherent multi-level medium with a tailored driving geometry that relies on two phase-mismatched standing-wave (SW) beams. This combination is essential for creating extra scattering channels over which, in addition, fully tunable optical nonreciprocal reflection can be attained. Our general approach is here adapted to four-level double-$\Lambda$ atoms that are found to exhibit distinct forms of nonreciprocal multi-channel scattering and quite sensitive to easily tunable parameters of two SW driving beams. The numerical results we present offer valuable insights into the field of non-Hermitian optical scattering and arise indeed from the interplay of interference among scattering processes and Bragg reflection.
\end{abstract}

\maketitle

\section{Introduction}
Investigations on optical nonreciprocity by trying to go beyond the Lorentz reciprocity~\cite{LRT} are of fundamental interest driven by the crucial role of nonreciprocal devices in applications like photonic signal processing and quantum networks~\cite{Nature453,Nat.P.7,PRL.125.123901,LSA.14.23}. It has been shown that a multitude of nonreciprocal devices, including unidirectional isolators~\cite{Nat.P.7}, circulators~\cite{Optica.5.279}, and amplifiers~\cite{PRA.109.023520}, can be designed via unusual techniques for achieving direction-dependent transmission properties. One intuitive technique relies on magneto-optical effects to break time reversal symmetry as the basis of Lorentz reciprocity, hence establishing optical nonreciprocity, with a magnetic field applied as an external bias~\cite{Nature.London.461,PRL.105.126804}. In view of the intractable difficulties in achieving low loss, small size, and less costly magnetic materials, however, magnet-free nonreciprocal devices have attracted growing interest and led to many frontier studies. Such devices may be realized by resorting to optomechanical interactions~\cite{Nat.Commun.7.13662,PRX.7.031001,PRA.109.043103,AQT}, spatiotemporal modulation~\cite{Nat.Photon.6,Nat.Photon.11,Nat.Photon.15,Nat. Commun.12.3746,PRL.128.173901}, stimulated Brillouin scattering~\cite{Nat.Cummun.6.6193,Nat.Phys.11.275,OE.23.25118}, spinning cavities~\cite{PRL.121.153601,Nature.558.569,PRA.111.013517}, moving atomic lattices~\cite{PRL.110.093901,PRL.110.223602,PRL.120.043901}, and thermal atomic vapors~\cite{PRL.125.123901,PRL.121.203602,Nat.Photon.12.744,PRL.123.033902}, all requiring external biases to break time reversal symmetry.

It is of particular interest that significant progress has been made recently in leveraging nonlinearities to achieve the transmission nonreciprocity~\cite{Nat.Photon.9.359,Nat.Photon.9.388,Nat.Electron.1.113}. However, much of this work relies on fixed optical structures and requires strong optical signals, thereby limiting the tunability and application scenarios of relevant nonreciprocal devices. Fortunately, such difficulties can be overcome by utilizing the effect of electromagnetically induced transparency (EIT) typically examined in the three-level $\Lambda$ atomic system~\cite{RMP.77.633}, which allows us to observe tunable nonlinear effects for weak optical signals by suppressing linear absorption on resonance. For instance, coherent four-wave mixing (FWM) in four-level double-$\Lambda$~\cite{PRA.60.4996,PRL.84.5308,PRA.70.053818,OL.35.3778,PRA.89.023839,PRA.93.033815,OE.24.1008} and double-ladder~\cite{PRA.82.053842,PRL.111.123602,OL.40.5674,OE.27.34611} atomic systems has been well investigated for a growing number of applications, including frequency conversion~\cite{OL.35.3778,PRA.89.023839}, squeezed light or biphoton generation~\cite{PRA.78.043816,PRA.84.053826,PRA.88.033845,PRL.94.183601,PRL.100.183603,PRA.106.023711,PRA.109.043711,PRA.110.063723}, optical storage or quantum memory~\cite{PRA.83.063823,PRA.86.053827,PRA.87.013845,NJP.16.113053}, and quantum information processing~\cite{Nature.457.859,Nature.478.360,OE.20.11057}. Returning to the topic of our concern, we notice that FWM in the double-$\Lambda$ atomic system has been explored for achieving an efficient transmission nonreciprocity~\cite{OE.30.6284} with advantages of dynamic tunability and flexible design, requiring no special structures and strong signals yet.

While devices like optical isolators necessitate breaking the Lorentz reciprocity theorem~\cite{Nat.P.7}, broader forms of nonreciprocity are clearly of interest as well, such as optical nonreciprocity in the reflection mode, also known as unidirectional invisibility and reflectionless~\cite{PRL.106.213901,PRA.87.012103,PRL.113.123004,PRA.105.043712,NJP.26.013048}. These nontrivial phenomena could be used for constructing photonic devices with more abundant isolation functions and have been achieved at the exceptional points exhibited by non-Hermitian structures, \textit{e.g.} those exhibiting parity-time ($\mathcal{PT}$) symmetry or antisymmetry~\cite{PRL.106.213901,PRA.87.012103,PRL.113.123004,PRA.105.043712,NJP.26.013048,PRA.91.033811,OL.48.5735,OE.24.4289,PRL.124.030401,PRL.129.153901}. The key strategy is to design phase-mismatched spatial modulations between real $n_{R}$ and imaginary $n_{I}$ parts of a complex refractive index~\cite{PRL.106.213901,PRA.87.012103,PRL.113.123004,PRA.105.043712,NJP.26.013048,PRA.91.033811,OL.48.5735} so that $\mathcal{PT}$ symmetry (antisymmetry) can be attained when $n_{R}$ and $n_{I}$ are even (odd) and odd (even) functions, respectively, in space. Again, we note that it is viable to realize $\mathcal{PT}$ symmetry or antisymmetry and hence similar nonreciprocal phenomena in linear EIT systems when two standing-wave (SW) fields are applied~\cite{PRL.113.123004,PRA.105.043712,NJP.26.013048,PRA.91.033811,OL.48.5735} to drive (or trap) atoms in an appropriate geometry. As far as we know, relevant studies have not been extended to nonlinear FWM systems because the problem is already complicated when atoms are driven by two travelling-wave (TW) fields.
\vspace{2mm}

Building upon these prominent proposals and experiments, in this work we suggest leveraging nonlinear wave-mixing to attain a fully tunable control of optical nonreciprocity in familiar multi-level coherent media. The essence is to combine the medium's intrinsic nonlinearities with a straightforward but effective driving that involves two phase-mismatched SW beams. Specifically, we consider a double-$\Lambda$ configuration based archetype where two right-arm transitions are driven by a coupling and a dressing SW field, out-of-phase with respect to one another, while two left-arm transitions serve as input channels for a probe and a signal field. The input probe and signal may impinge either in the same or in the opposite direction [see Fig.~\ref{fig1}(a)], hence the current archetype design implements a general \textit{four-mode} input-port device, \textit{viz}. two separate frequency modes and two different spatial modes [see Fig.~\ref{fig1}(b)]. Here, we characterize optical nonreciprocity in two reflection and two transmission channels. Distinct forms of nonreciprocity are found that entail both direct (same-color) and cross (different-color) reflections and likewise for cross transmission though direct transmission remains reciprocal yet, corresponding thus to a \textit{four-channel} scattering scheme. This may be understood as resulting from the interplay between cross and direct non-Hermitian scattering, also enhanced by well-established Bragg reflection conditions.

Numerical results will be discussed for a cold sample of $^{87}$Rb atoms as the scattering medium and restricted to the case of a ``single-mode" input with only a probe or a signal entering our four-mode atomic device. Accurate tuning and optimization of the nonreciprocal response are achieved through a careful engineering of the outgoing reflection and transmission amplitudes, which turn out to be sensitive to easily adjustable parameters of two strong SW driving beams (\textit{e.g.} detuning, strength, and phase mismatch), making our proposal viable for a potential implementation. Our findings address, at least in part, the limited tunability of scattering observed in earlier approaches that relied on fixed device architectures. They also lay the ground for achieving multi-color optical nonreciprocity within a single device. Although the scheme we propose is general, our results for an atomic implementation further aim to provide insights into the important field of non-Hermitian optical scattering.

\section{\label{sec2}Model and Equations}

\begin{figure}[ptbh]
\centering
\includegraphics[width=0.48\textwidth]{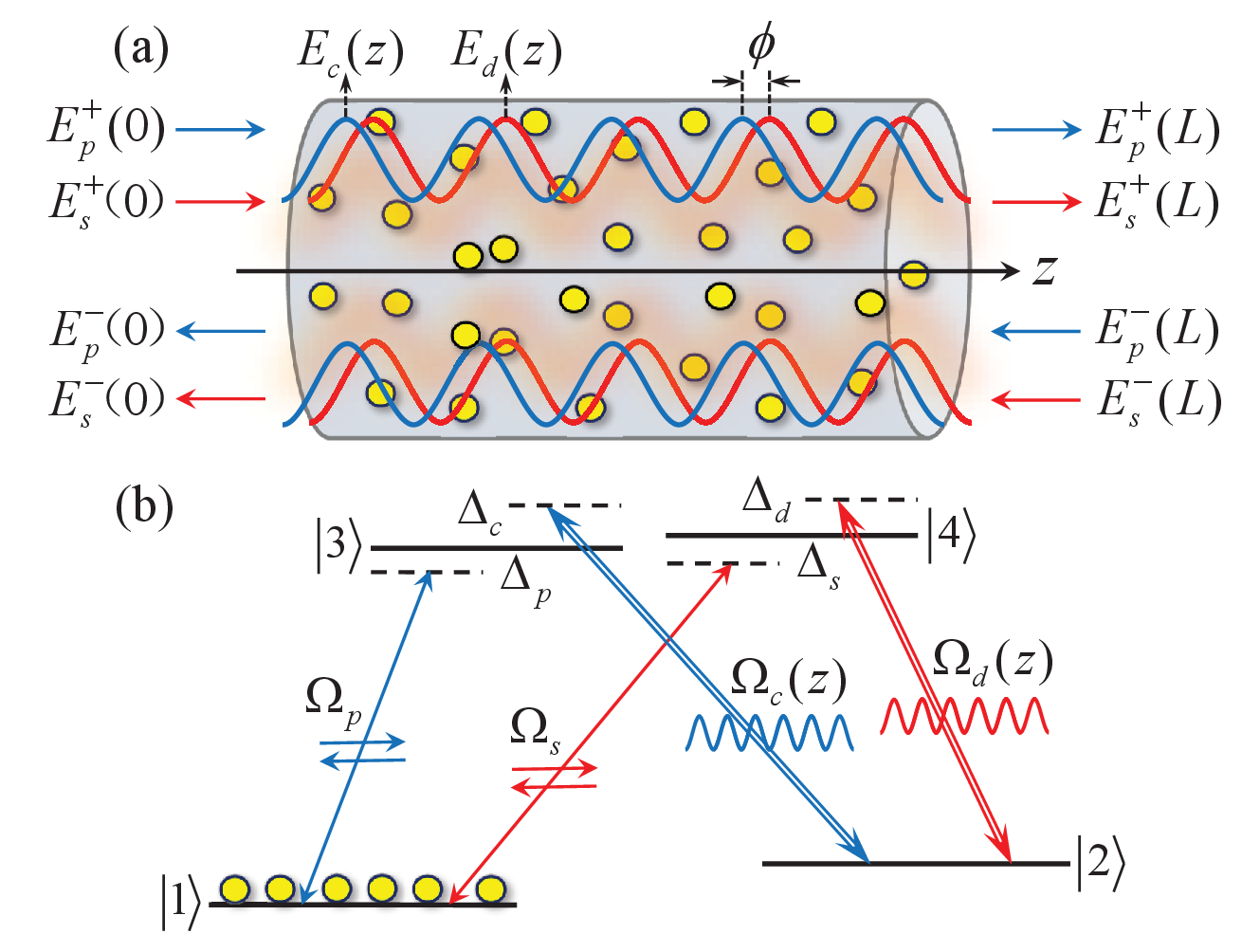}
\caption{(a) Schematic diagram of a four-mode four-channel atomic device driven by a coupling $E_{c}(z)$ field and a dressing $E_{d}(z)$ field in the form of phase-mismatched ($\phi$) SW gratings. A probe $E_{p}^{+}(0)$ or signal $E_{s}^{+}(0)$ input field (left) scatters into a collection of output fields $\{E_{p}^{-}(0),E_{s}^{-}(0),E_{p}^{+}(L),E_{s}^{+}(L)\}$ through four channels, leveraging third-order nonlinearities of the atomic device and Bragg reflections of two SW gratings. Likewise for a probe $E_{p}^{-}(L)$ or signal $E_{s}^{-}(L)$ input field entering the atomic device from the opposite side (right). Here $+$ ($-$) denotes the forward (backward) $z$ ($-z$) propagation. (b) Level configuration of a double-$\Lambda$ system for all atoms (yellow filled circles) in the scattering device driven by two strong fields (thick transition lines) of Rabi frequencies $\Omega_{c,d}(z)$ and detunings $\Delta_{c,d}$ together with two weak fields (thin transition lines) of Rabi frequencies $\Omega_{p,s}$ and detunings $\Delta_{p,s}$.}
\label{fig1}
\end{figure}

We aim to investigate the four-mode four-channel outgoing states resulted from the non-Hermitian nonlinear scattering in a coherently driven atomic device of length $L$ in the simple case of a single-mode input field. Besides the usual transmission and reflection channels, two additional effective scattering channels can be constructed by leveraging (i) the intrinsic optical nonlinearities of this atomic device and (ii) the tunable SW geometries of two driving fields. The schematic illustration of such an enlarged or extended scattering-channel concept is depicted in Fig.~\ref{fig1}(a) where a forward probe field $E_{p}^{+}(0)$, \textit{e.g.} enters from the left input port at $z=0$ and scatters into not only $E_{p}^{+}(L)$ (transmission) and $E_{p}^{-}(0)$ (reflection) through two usual channels but also $E_{s}^{+}(L)$ (transmission) and $E_{s}^{-}(0)$ (reflection) through other two channels at a different signal frequency. The two additional channels are foreseen here to emerge from distinct mechanisms: the first (cross transmission) simply from nonlinear frequency conversion, whereas the second (cross reflection) from a combination of the atomic device's nonlinearities with two driving fields' SW geometries. The same holds for a forward signal field $E_{s}^{+}(0)$ entering also from the left input port at $z=0$. Similarly, a backward probe $E_{p}^{-}(L)$ or signal $E_{s}^{-}(L)$ field entering from the right input port at $z=L$ scatters again into a four-mode four-channel outgoing state including the transmitted pair $\{E_{p}^{-}(0),E_{s}^{-}(0)\}$ and the reflected pair $\{E_{p}^{+}(L),E_{s}^{+}(L)\}$.

The four scattering channels leading to a collection of the outgoing fields $\{E_{p}^{-}(0),E_{s}^{-}(0),E_{p}^{+}(L),E_{s}^{+}(L)\}$ are characterized by four always existing complex amplitudes for the direct (or \textit{same-color}) transmission
\begin{eqnarray}\label{equ1}
t_{pp}^{++} &=& \frac{E_{p}^{+}(L)}{E_{p}^{+}(0)},\,\ t_{ss}^{++}=\frac{E_{s}^{+}(L)}{E_{s}^{+}(0)}, \nonumber \\
t_{pp}^{--} &=& \frac{E_{p}^{-}(0)}{E_{p}^{-}(L)},\,\ t_{ss}^{--}=\frac{E_{s}^{-}(0)}{E_{s}^{-}(L)},
\end{eqnarray}
and four usually missing complex amplitudes for the cross (or \textit{different-color}) transmission
\begin{eqnarray}\label{equ2}
t_{ps}^{++} &=& \frac{E_{s}^{+}(L)}{E_{p}^{+}(0)},\,\ t_{sp}^{++}=\frac{E_{p}^{+}(L)}{E_{s}^{+}(0)}, \nonumber \\
t_{ps}^{--} &=& \frac{E_{s}^{-}(0)}{E_{p}^{-}(L)},\,\ t_{sp}^{--}=\frac{E_{p}^{-}(0)}{E_{s}^{-}(L)},
\end{eqnarray}
originating solely from nonlinear frequency conversion along the $\pm z$ directions. Similarly, one has
\begin{eqnarray}\label{equ3}
r_{pp}^{+-} &=& \frac{E_{p}^{-}(0)}{E_{p}^{+}(0)},\,\ r_{ss}^{+-}=\frac{E_{s}^{-}(0)}{E_{s}^{+}(0)}, \nonumber \\
r_{pp}^{-+} &=& \frac{E_{p}^{+}(L)}{E_{p}^{-}(L)},\,\ r_{ss}^{-+}=\frac{E_{s}^{+}(L)}{E_{s}^{-}(L)},
\end{eqnarray}
for the direct (or \textit{same-color}) reflection and
\begin{eqnarray}\label{equ4}
r_{ps}^{+-} &=& \frac{E_{s}^{-}(0)}{E_{p}^{+}(0)},\,\ r_{sp}^{+-}=\frac{E_{p}^{-}(0)}{E_{s}^{+}(0)}, \nonumber \\
r_{ps}^{-+} &=& \frac{E_{s}^{+}(L)}{E_{p}^{-}(L)},\,\ r_{sp}^{-+}=\frac{E_{p}^{+}(L)}{E_{s}^{-}(L)},
\end{eqnarray}
for the cross (or \textit{different-color}) reflection resulted from the interplay of nonlinear frequency conversion and Bragg reflection. Keep in mind that all $16$ reflection and transmission amplitudes are defined here for a `single-mode' input field. In this case, when $E_{p}^{+}(0)\ne 0$ and all other denominators are zero, for instance, the resulting transmission and reflection amplitudes correspond to those in the upper-left grid corners in Eqs.~(\ref{equ1})--(\ref{equ4}). Similarly, we just need to consider the transmission and reflection amplitudes in the upper-right, lower-left, or lower-right grid corners in Eqs.~(\ref{equ1})--(\ref{equ4}) when only $E_{s}^{+}(0)\ne 0$, $E_{p}^{-}(L)\ne 0$, or $E_{s}^{-}(L)\ne 0$ without other input fields.
\vspace{2mm}

The four outgoing electric fields entering above complex amplitudes can be obtained by solving steady-state Maxwell equations in the slowly-varying envelope approximation (see Eq.~(\ref{B1}) in Appendix~\ref{appB}) with polarizations $P_{31}(z)$ and $P_{41}(z)$ oscillating, respectively, at the probe and signal frequencies as driving terms. A spatial integration of the resultant four-mode coupled equations (see Eq.~(\ref{B3}) in Appendix~\ref{appB}) would yield the spatial variations of $\{E_{p}^{+}(z),E_{p}^{-}(z),E_{s}^{+}(z),E_{s}^{-}(z)\}$ in the atomic device. Then, taking $z=0$ and $z=L$ for two device boundaries, it is straightforward to further obtain
\begin{equation}\label{equ5}
\left( {\begin{array}{*{20}{c}}
E_{p}^{+}(L)\\
E_{p}^{-}(L)\\
E_{s}^{+}(L)\\
E_{s}^{-}(L)
\end{array}} \right) =
e^{\hat{X}}
\left( {\begin{array}{*{20}{c}}
E_{p}^{+}(0)\\
E_{p}^{-}(0)\\
E_{s}^{+}(0)\\
E_{s}^{-}(0)
\end{array}} \right)\\
\equiv
\hat{M}\left( {\begin{array}{*{20}{c}}
E_{p}^{+}(0)\\
E_{p}^{-}(0)\\
E_{s}^{+}(0)\\
E_{s}^{-}(0)
\end{array}} \right),\\
\end{equation}
which connects four electric fields on the left ($z=0$) to those on the right ($z=L$) via a $4\times4$ transfer matrix $\hat{M}$.
The matrix elements $M_{ij}$, when computed for appropriate boundary conditions~\cite{bc}, enable us to determine the transmission and reflection amplitudes in Eqs.~(\ref{equ1})--(\ref{equ4}). Detailed general expressions of these amplitudes in terms of various $M_{ij}$ for two SW driving fields can be found in Appendix~\ref{appB}, while the usual case with two TW driving fields where only transmission amplitudes are relevant is first discussed in Appendix~\ref{appA} for comparison.
\vspace{2mm}

We will now apply our previous general discussions to the specific scenario involving two spatially \textit{periodic} non-Hermitian polarizations $P_{31}(z)$ and $P_{41}(z)$. This can be realized by using a coupling and a dressing fields whose electric fields in the SW pattern are given by
\begin{eqnarray}\label{equ6}
E_{c}(z) &=& 2\mathcal{E}_{0}\cos(k_{c}z), \nonumber \\
E_{d}(z) &=& 2\mathcal{E}_{0}\cos(k_{d}z+\phi),
\end{eqnarray}
with a phase mismatch ($\phi$) but equal maximal strengths ($2\mathcal{E}_{0}$). Here, $k_{c}=2\pi\cos\theta_{c}/\lambda_{c}$ and $k_{d}=2\pi\cos\theta_{d}/\lambda_{d}$ are introduced to describe different wavenumbers of the two SW fields, with $\lambda_{c}$ and $\lambda_{d}$ ($\theta_{c}$ and $\theta_{d}$) representing their wavelengths (misalignments with respect to the $\pm z$ directions). The required two polarizations can be evaluated then by considering a four-level atomic system in the double-$\Lambda$ configuration as shown by Fig.~\ref{fig1}(b), where the \emph{weak} probe $E_{p}$ and signal $E_{s}$ fields of frequencies $\omega_{p}$ and $\omega_{s}$ drive, respectively, two left-arm $|1\rangle\leftrightarrow|3\rangle $ and $|1\rangle\leftrightarrow|4\rangle$ transitions while the \emph{strong} coupling $E_{c}(z)$ and dressing $E_{d}(z)$ fields of frequencies $\omega_{c}$ and $\omega_{d}$ drive, respectively, two right-arm $|2\rangle\leftrightarrow|3\rangle $ and $|2\rangle\leftrightarrow|4\rangle$ transitions. A complete description of such coherent atom-light interactions further relies on four detunings defined by $\Delta_{p}=\omega_{p}-\omega_{31}$, $\Delta_{s}=\omega_{s}-\omega_{41}$, $\Delta_{c}=\omega_{c}-\omega_{32}$, and $\Delta_{d}=\omega_{d}-\omega_{42}$ as well as four Rabi frequencies defined by $\Omega_{p}=E_{p}d_{13}/2\hbar$, $\Omega_{s}=E_{s}d_{14}/2\hbar$, $\Omega_{c}(z)= E_{c}(z)d_{23}/2\hbar$, and $\Omega_{d}(z)=E_{d}(z)d_{24}/2\hbar$, with $\omega_{ji}$ ($d_{ij}$) being resonant frequencies (dipole moments) of relevant transitions.

Under both electric-dipole and rotating-wave approximations, we write down the interaction Hamiltonian
\begin{eqnarray}\label{eq2}
H_{I}& =& -\hbar[\delta_{2}|2\rangle\langle2|+\Delta_{p}|3\rangle\langle3|+\delta_{3}|4\rangle\langle4|]-\hbar
[\Omega_{p}|3\rangle\langle1|\nonumber \\
&+& \Omega_{s}|4\rangle\langle1|+\Omega_{c}(z)|3\rangle\langle2|+\Omega_{d}(z)|4\rangle\langle2|+h.c.],
\end{eqnarray}
where $\delta_{2}=\Delta_{p}-\Delta_{c}$ and $\delta_{3}=\Delta_{p}-\Delta_{c}+\Delta_{d}$ are two-photon and three-photon detunings, respectively.
Above, we have considered $\Delta_{s}=\delta_{3}$ to ensure energy conservation in a desired FWM process. This Hamiltonian can be employed together with the Lindblad superoperator $\mathcal{L}(\rho)$ accounting for population decay rates $\Gamma_{ij}$ and coherence dephasing rates $\gamma_{ij}$ to derive a set of dynamic equations for $16$ density matrix elements $\rho_{ij}$.
Setting $\partial_{t}\rho_{ij}=0$ in the limit of weak probe and signal fields ($\Omega_{p,s}\to 0$), it is not difficult to obtain two steady-state solutions
\begin{widetext}
\begin{eqnarray}
\label{equ8}
\rho_{31}^{\langle1\rangle}(z)&=&i\frac{[g_{21}g_{41}+\Omega_{d}^{2}(z)]\Omega_{p}-\Omega_{c}(z)\Omega_{d}(z)\Omega_{s}}
{g_{31} [{g_{21}g_{41}+\Omega_{d}^{2}(z)]+g_{41} \Omega_{c}^{2}(z)}}=A(z)\Omega_{p}+B(z)\Omega_{s}, \\
\rho_{41}^{\langle1\rangle}(z)&=&i\frac{[g_{21}g_{31}+\Omega_{c}^{2}(z)]\Omega_{s}-\Omega_{c}(z)\Omega_{d}(z)\Omega_{p}}
{g_{31} [{g_{21}g_{41}+\Omega_{d}^{2}(z)]+g_{41} \Omega_{c}^{2}(z)}}=C(z)\Omega_{s}+D(z)\Omega_{p}, \notag
\end{eqnarray}
\end{widetext}
while $\rho_{11}^{\langle1\rangle}(z)\to1$ to the first order of $\Omega_{p}$ and $\Omega_{s}$ with all other solutions being totally negligible. Here, we have introduced the complex dephasing rates $g_{21}=\gamma_{21}-i\delta_{2}$, $g_{31}=\gamma_{31}-i\Delta_{p}$, and $g_{41}=\gamma_{41}-i\delta_{3}$ where $\gamma_{21}$ (dipole forbidden transition) is at least three-order smaller than $\gamma_{31}\simeq\gamma_{41}$ (dipole allowed transitions). The out-of-phase spatial periodicities of $\rho_{31}^{\langle1\rangle}(z)$ and $\rho_{41}^{\langle1\rangle}(z)$ can be seen in a much clearer way by further considering
\begin{eqnarray}
\label{equ9}
\Omega_{c}^{2}(z) &=& 2G_{0}^{2}[1+\cos(2k_{0}z)], \notag \\
\Omega_{d}^{2}(z) &=& 2G_{0}^{2}[1+\cos(2k_{0}z+2\phi)],\\
\Omega_{c}(z)\Omega_{d}(z) &=& 2G_{0}^{2}[\cos(2k_{0}z+\phi)+\cos(\phi)], \notag
\end{eqnarray}
where we have defined $G_{0}=\mathcal{E}_{0}d_{0}/2\hbar$. We recall that in this equation (i) $d_{23}=d_{24}=d_{0}$ and (ii) $k_{c}=k_{d}=k_{0}$
have been taken for simplicity. The former can be easily achieved via a proper choice of relevant atomic states as we do in the next section. The latter, critical for ensuring a common Bragg condition for both probe and signal fields so that their reflections can be simultaneously enhanced, is viable by carefully adjusting $\theta_{c}$ and $\theta_{d}$.

Note in particular that $\rho_{31}^{\langle1\rangle}(z)$ reduces to the familiar expression $i\Omega_{p}/g_{31}$ for a two-level coherence when both coupling and dressing fields are absent ($\Omega_{c,d}(z)\to 0$) or the coherence $ig_{21}\Omega_{p}/[g_{21}g_{31}+\Omega^{2}_{c}(z)]$ for a three-level $\Lambda$ system when only the dressing field vanishes ($\Omega_{d}(z)\to0$). Subtle is instead the effect of the SW dressing field on the probe response, manifesting in two distinct ways. First, it renormalizes the two-level coherence through the extra spatial modulation term proportional to $\Omega_{d}^{2}(z)$ inside the two square brackets of Eq.~(\ref{equ8}). Second, it quantifies a nonlinear effect of the signal strength $\Omega_{s}$ on the probe response through the cross-modulation term proportional to $\Omega_{c}(z)\Omega_{d}(z)$ in the numerator of Eq.~(\ref{equ8}). The same holds for $\rho_{41}^{\langle1\rangle}(z)$ with an exchange of the roles played by the probe and signal fields. We have purposely reformulated Eq.~(\ref{equ8}) to highlight the probe and signal coupled \textit{propagation} dynamics using the last terms on the right-hand side of above expressions for $\rho_{31}^{\langle1\rangle}(z)$ and $\rho_{41}^{\langle1\rangle}(z)$. This coupling is quantified by the cross terms $B(z)$ and $D(z)$, both of which are proportional to $\Omega_{c}(z)\Omega_{d}(z)$ - the product of two SW gratings. These cross terms are not only crucial for evaluating the mutual influence between probe and signal modes and their combined propagation dynamics, but also enable to modulate the nonlinear mixing process in space for generating multiple outgoing scattering channels, as detailed in the next section.

With above results for $\rho_{31}^{\langle1\rangle}(z)$ and $\rho_{41}^{\langle1\rangle}(z)$, it is straightforward to write down the following polarizations
\begin{eqnarray}
\label{equ10}
P_{31}(z) &=& Nd_{13}\rho_{31}^{\langle1\rangle}(z)=Nd_{13}[A(z)\Omega_{p}+B(z)\Omega_{s}] \notag\\
&\equiv& P_{31}^{(l)}(z)+P_{31}^{(n)}(z), \notag\\
P_{41}(z) &=& Nd_{14}\rho_{41}^{\langle1\rangle}(z)=Nd_{14}[C(z)\Omega_{s}+D(z)\Omega_{p}] \notag\\
&\equiv& P_{41}^{(l)}(z)+P_{41}^{(n)}(z),
\end{eqnarray}
for a cold atomic sample of density $N$. Here, $P_{31}(z)$ has been split into the ``linear" (direct) $P_{31}^{(l)}(z)$ and ``nonlinear" (cross) $P_{31}^{(n)}(z)$ components, which jointly determine the propagation dynamics of the probe field (see also Eqs.~(\ref{equ11}) and (\ref{equ12}) below). Similarly, $P_{41}^{(l)}(z)$ and $P_{41}^{(n)}(z)$ represent the linear and nonlinear responses, respectively, at the signal frequency. Above discussions indicate the potential for an all optically controlled scheme that combines two frequencies ($\omega_{p,s}$) and two directions ($\pm z$) to operate an archetype of four-mode four-channel scattering devices. Needless to say, operations involving more than four modes are also achievable, naturally enabling a larger number of outgoing scattering channels within the same atomic sample driven by two SW gratings. This is possible provided an appropriate configuration of atomic levels along with their symmetries is selected.

\begin{figure}[t]
\begin{centering}
\includegraphics[width=0.48\textwidth]{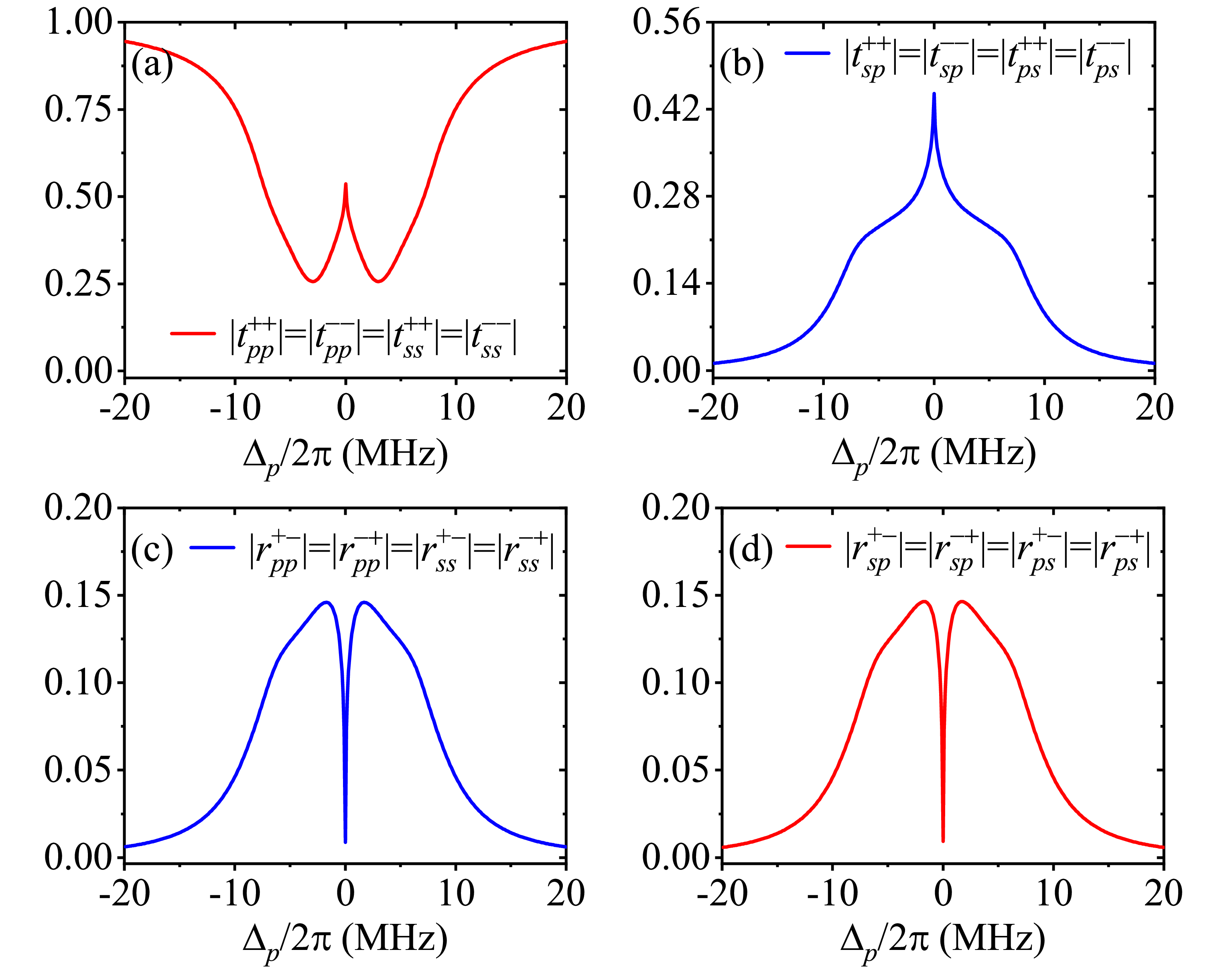}
\caption{Moduli of direct (a) and cross (b) transmission amplitudes as well as direct (c) and cross (d) reflection amplitudes vs probe detuning $\Delta_{p}$. The two SW driving fields are resonant and balanced with $\Delta_{c}=\Delta_{d}=0$, $G_{0}=2\pi\times7.0$ MHz, and $\phi=0$. The scattering medium  of cold $^{87}{\rm{Rb}}$ atoms exhibits density $N=1.0\times10^{12}$ cm$^{-3}$, length $L=0.4$ mm, dephasing rates $\gamma_{31}=\gamma_{41}=10^{3}\gamma_{21}=2\pi\times 3.0$ MHz, and dipole moments $d_{23,24}=\sqrt{3}d_{13,14}=1.268\times10^{-29}$ Cm.}
\label{fig2}
\end{centering}
\end{figure}

\section{\label{sec3}Results and Discussion}
In this section, we aim at exploring the nonreciprocal scattering effects that originate from coherent nonlinear mixing (FWM) attained with out-of-phase periodic coupling and dressing fields (SW). We specifically consider \textit{four} `single-mode' input cases whereby a probe or a signal field enters the four-mode four-channel atomic sample in Fig.~\ref{fig1}(a) from either left or right side. In each of the four cases, the four-channel outgoing scattering states are assessed by calculating the transmission and reflection amplitudes in Eq.~(\ref{equ1})--Eq.~(\ref{equ4}). For numerical calculations, we focus on a scattering medium consisting of cold $^{87}{\rm{Rb}}$ atoms with the four levels $|1\rangle\equiv |5^{2}S_{1/2},F=1,m=-1\rangle$, $|2\rangle\equiv |5^{2}S_{1/2},F=2,m=1\rangle $, $|3\rangle\equiv |5^{2}P_{1/2},F=1,m=0\rangle $, and $|4\rangle\equiv |5^{2}P_{1/2},F=2,m=0\rangle$ chosen on the D1 line exhibiting the wavelengths $\lambda_{p,s,c,d} \simeq 795$ nm.

\begin{figure}[t]
\includegraphics[width=0.48\textwidth]{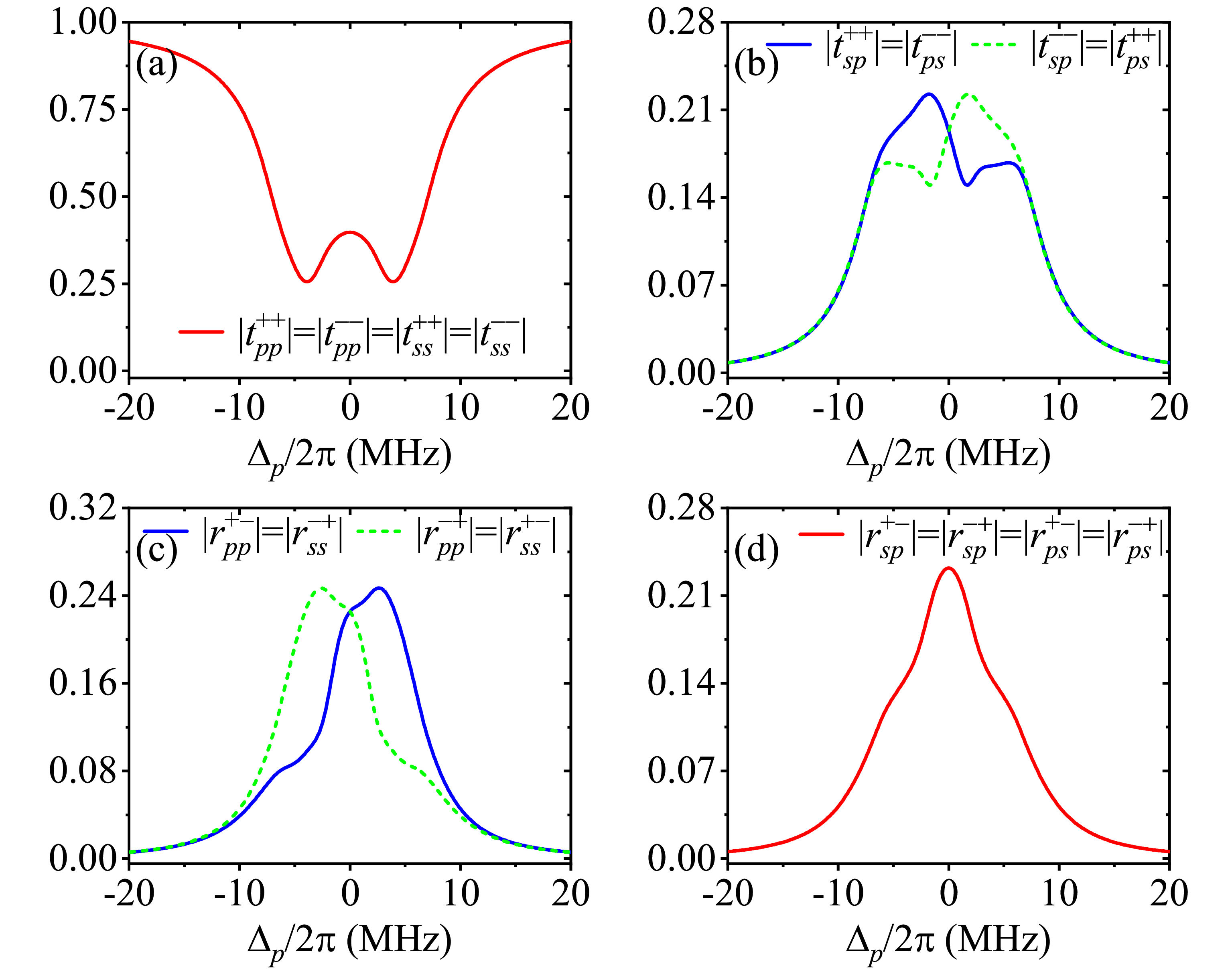}
\caption{Moduli of direct (a) and cross (b) transmission amplitudes as well as direct (c) and cross (d) reflection amplitudes vs probe detuning $\Delta_{p}$ attained with the same parameters as in Fig.~\ref{fig2} except a nonzero phase shift $\phi=\pi/4$.}\label{fig3}
\end{figure}

We first plot in Fig.~\ref{fig2} the moduli of all transmission and reflection amplitudes against probe detuning $\Delta_{p}$ by taking $\phi=0$ to make the SW coupling and dressing fields spatially modulated in phase. It is easy to observe that the eight transmission amplitudes remain reciprocal with $|t_{pp,ss}^{++}|=|t_{pp,ss}^{--}|$ and $|t_{ps,sp}^{++}|=|t_{ps,sp}^{--}|$, an invariance upon the exchange of input directions `$+\leftrightarrow -$', like what are shown in Appendix~\ref{appA} for the TW coupling and dressing fields. The main difference lies in that here we have also eight non-vanishing reflection amplitudes except when $|\Delta_{p}|$ is too large or tends to zero, which are equally reciprocal with $|r_{pp,ss}^{+-}|=|r_{pp,ss}^{-+}|$ and $|r_{ps,sp}^{+-}|=|r_{ps,sp}^{-+}|$. Moreover, we should note that $|t_{pp}^{\pm\pm}|=|t_{ss}^{\pm\pm}|$, $|t_{ps}^{\pm\pm}|=|t_{sp}^{\pm\pm}|$, $|r_{pp}^{\pm\mp}|=|r_{ss}^{\pm\mp}|$, and $|r_{ps}^{\pm\mp}|=|r_{sp}^{\pm\mp}|$, which indicate another invariance upon the exchange of input fields `$p\leftrightarrow s$' as a result of the symmetric driving detunings ($\Delta_{c}=\Delta_{d}$). It is of no doubt that this invariance will be broken for the asymmetric driving detunings ($\Delta_{c}\ne\Delta_{d}$).

Then we examine in Fig.~\ref{fig3} another case with $\phi=\pi/4$ instead to make the SW coupling and dressing fields spatially modulated out of phase. It is interesting that nonreciprocal behaviors occur for both cross transmissions with $|t_{ps,sp}^{++}|\ne|t_{ps,sp}^{--}|$ and direct reflections with $|r_{pp,ss}^{+-}|\ne|r_{pp,ss}^{-+}|$ when $|\Delta_{p}|$ is neither vanishing nor too large. But, reciprocal direct transmissions with $|t_{pp,ss}^{++}|=|t_{pp,ss}^{--}|$ and cross reflections with $|r_{ps,sp}^{+-}|=|r_{ps,sp}^{-+}|$ remain valid everywhere, \textit{i.e.} independent of $\Delta_{p}$. Moreover, we have an invariance upon the simultaneous exchange of input fields and directions with $|t_{pp,ss}^{++}|=|t_{ss,pp}^{--}|$, $|t_{ps,sp}^{++}|=|t_{sp,ps}^{--}|$, $|r_{pp,ss}^{+-}|=|r_{ss,pp}^{-+}|$, and $|r_{ps,sp}^{+-}|=|r_{sp,ps}^{-+}|$. That means, the phase shift $\phi$ alone cannot break all intrinsic symmetries in our double-$\Lambda$ atomic system, though it plays a crucial role in achieving nonreciprocal cross transmissions and direct reflections. It is also easy to learn from Fig.~\ref{fig3} an invariance of all transmission and reflection amplitudes upon the simultaneous exchange of input directions and detuning signs, so it is enough to focus just on the regime of $\Delta_{p}\ge0$ in following discussions for simplicity.

\begin{figure}
\begin{centering}
\includegraphics[width=0.48\textwidth]{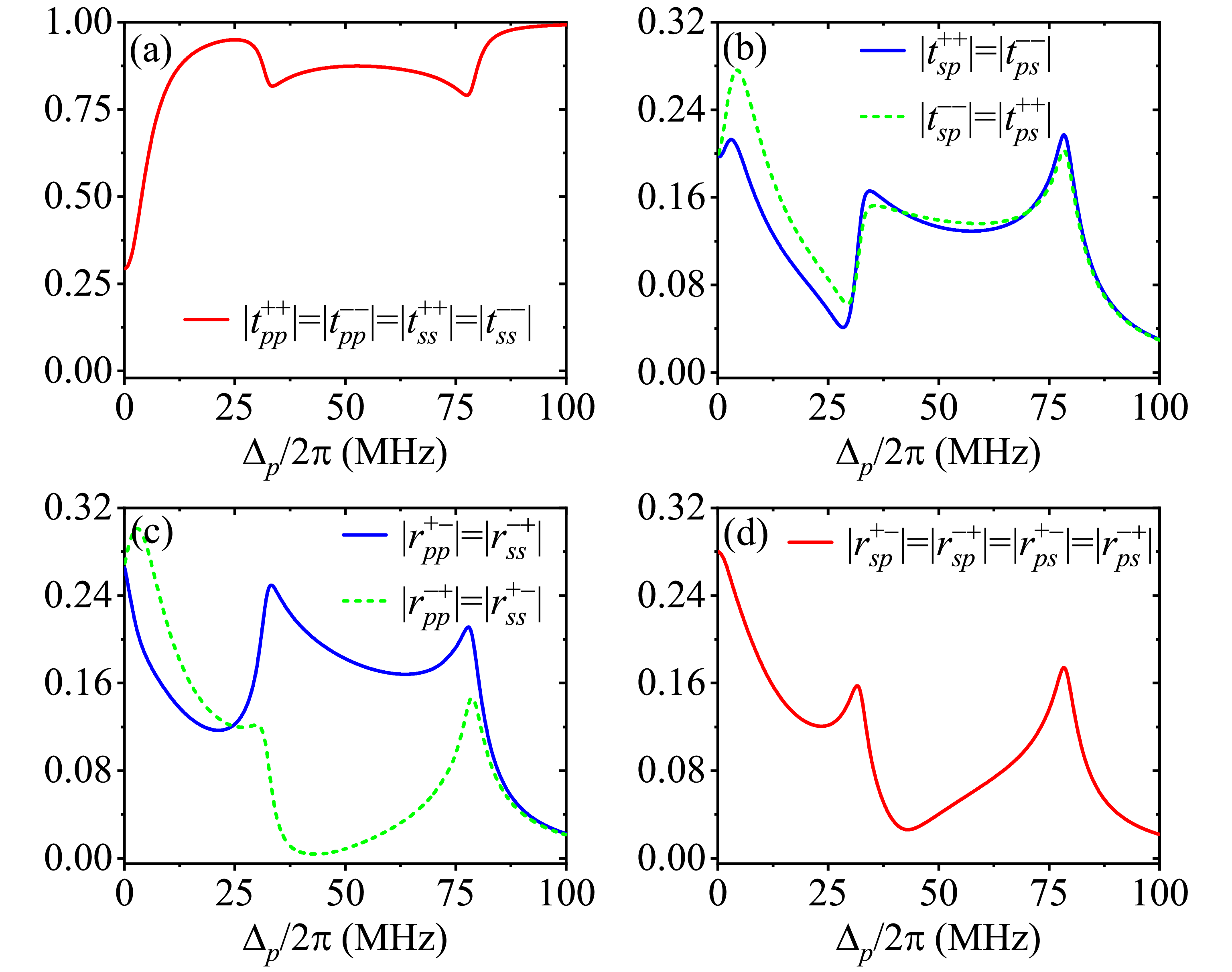}
\caption{Moduli of direct (a) and cross (b) transmission amplitudes as well as direct (c) and cross (d) reflection amplitudes vs probe detuning $\Delta_{p}$ attained with the same parameters as in Fig.~\ref{fig3} except $G_{0}=2\pi\times85$ MHz and $L=0.6$ mm.}\label{fig4}
\end{centering}
\end{figure}

\begin{figure*}
\begin{centering}
\includegraphics[width=0.96\textwidth]{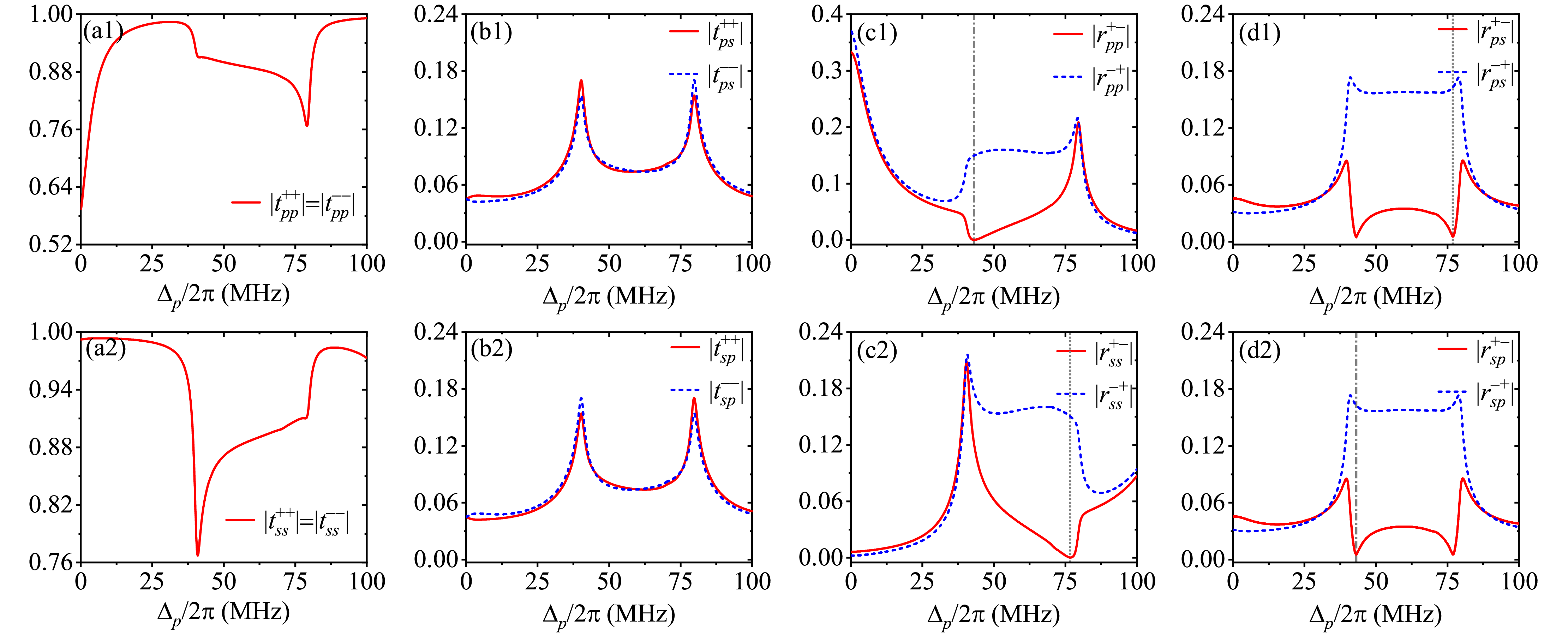}
\caption{Moduli of direct (a) and cross (b) transmission amplitudes as well as direct (c) and cross (d) reflection amplitudes vs probe detuning $\Delta_{p}$ attained with the same parameters as in Fig.~\ref{fig4} except $\Delta_{c}=-\Delta_{d}=2\pi\times60$ MHz and $G_{0}=2\pi\times70$ MHz.}\label{fig5}
\end{centering}
\end{figure*}

We have numerically checked that all reflections gradually increase until becoming saturated, direct transmissions continuously decrease until approaching zero, while cross transmissions first increase and then decrease, for a longer and longer atomic sample when the two driving fields are kept strong enough. In light of this fact, we have increased $L$ from $0.4$ mm to $0.6$ mm and $G_{0}/2\pi$ from $7.0$ MHz to $85$ MHz for seeking more favorable scattering results in Fig.~\ref{fig4}. It is clear that nonreciprocal direct reflections further turn out to be unidirectional direct reflections in a visible region of $\Delta_{p}>0$ though cross transmissions remain to be just nonreciprocal. To be more specific, we have found $|r_{pp}^{-+}|=|r_{ss}^{+-}|\to0$ around $\Delta_{p}/2\pi\simeq40$ MHz where $|r_{pp}^{+-}|=|r_{ss}^{-+}|$ are nonzero instead and up to $0.2$ in a broader region of $\Delta_{p}$. This interesting behavior of unidirectional direct reflection could even be reversed with $|r_{pp}^{+-}|=|r_{ss}^{-+}|\to 0$ around $\Delta_{p}/2\pi\simeq-40$ MHz due to the exchange symmetry of $|r_{pp,ss}^{+-}(\Delta_{p})|=|r_{pp,ss}^{-+}(-\Delta_{p})|$ with respect to two detuning signs. Note, however, that the two symmetric detuning regions of unidirectional direct reflection will gradually shrink toward $\Delta_{p}=0$ and finally disappear if we reduce $G_{0}$ as in Fig.~\ref{fig3}.

The above results refer to the specific case of resonant coupling and dressing fields with $\Delta_{c}=\Delta_{d}=0$. Hence, we examine in Fig.~\ref{fig5} a more general case where the coupling and dressing fields exhibit opposite nonzero detunings $\Delta_{c}=-\Delta_{d}\ne0$. We can see that direct reflections remain nonreciprocal with $|r_{pp}^{+-}|\ne|r_{pp}^{-+}|$ and $|r_{ss}^{+-}|\ne|r_{ss}^{-+}|$ in a quite wide region, but are unidirectional just at a specific point of $\Delta_{p}/2\pi\simeq43$ MHz for the probe field or $\Delta_{p}/2\pi\simeq76$ MHz for the signal field. It is more interesting that cross reflections exhibit similar nonreciprocal behaviors with $|r_{ps}^{+-}|\ne|r_{ps}^{-+}|$ and $|r_{sp}^{+-}|\ne|r_{sp}^{-+}|$, which are unidirectional at both $\Delta_{p}/2\pi\simeq43$ MHz and $\Delta_{p}/2\pi\simeq76$ MHz identical to direct probe and signal reflections, respectively. That means, there exists no reflected probe (signal) field leaving from the left side due to $|r_{pp}^{+-}|=|r_{sp}^{+-}|=0$ ($|r_{ss}^{+-}|=|r_{ps}^{+-}|=0$) at $\Delta_{p}/2\pi\simeq43$ MHz ($76$ MHz) no matter a probe or a signal field is input. Note also that cross transmissions remain nonreciprocal especially around $\Delta_{p}/2\pi\simeq43$ MHz and $76$ MHz, albeit in a less evident way. Moreover, the simultaneous exchanges of two input fields and directions cannot result in the invariance of $|t_{pp,ss}^{++}|=|t_{ss,pp}^{--}|$, $|t_{ps,sp}^{++}|=|t_{sp,ps}^{--}|$, $|r_{pp,ss}^{+-}|=|r_{ss,pp}^{-+}|$, and $|r_{ps,sp}^{+-}|=|r_{sp,ps}^{-+}|$ again.

The nonreciprocal and unidirectional scattering behaviors we find can be explained by extending the language of non-Hermitian optics~\cite{PRL.106.213901,PRA.87.012103,PRL.113.123004,PRA.105.043712,NJP.26.013048,PRA.91.033811,OL.48.5735} from the well known linear response case to the FWM regime. As noted above, our atomic sample driven by the strong SW coupling and dressing fields can be seen as an all-optical scattering device with four modes corresponding to different choices of frequencies $\omega_{p,s}$ and directions $\pm z$ of the weak probe and signal fields. Two polarizations governing the scattering processes as given by Eq.~(\ref{equ10}) can be described in terms of following response functions for the probe and signal fields: the {\it direct} (or linear) susceptibilities
\begin{eqnarray}\label{equ11}
\chi_{p}^{(l)}(z) &=& \frac{P_{31}^{(l)}(z)}{\varepsilon_{0}E_{p}}=\frac{Nd_{13}^{2}A(z)}{2\varepsilon_{0}\hbar}, \nonumber\\
\chi_{s}^{(l)}(z) &=& \frac{P_{41}^{(l)}(z)}{\varepsilon_{0}E_{s}}=\frac{Nd_{14}^{2}C(z)}{2\varepsilon_{0}\hbar},
\end{eqnarray}
and the {\it cross} (or nonlinear) susceptibilities
\begin{eqnarray}\label{equ12}
\chi_{p}^{(n)}(z) &=& \frac{P_{31}^{(n)}(z)}{2\varepsilon_{0}\mathcal{E}_{0}^{2}E_s}
=\frac{Nd_{13}d_{14}B(z)}{4\varepsilon_{0}\hbar\mathcal{E}_{0}^{2}}, \nonumber\\
\chi_{s}^{(n)}(z) &=& \frac{P_{41}^{(n)}(z)}{2\varepsilon_{0}\mathcal{E}_{0}^{2}E_p}
=\frac{Nd_{13}d_{14}D(z)}{4\varepsilon_{0}\hbar\mathcal{E}_{0}^{2}}.
\end{eqnarray}
Clearly, linear susceptibilities $\chi_{p,s}^{(l)}$ are the only ones that would survive even if we switch off the dressing and coupling fields; in nonlinear susceptibilities $\chi_{p,s}^{(n)}$ the main dependence on the dressing and coupling fields has been factored out. We note, however, that both $\chi_{p,s}^{(l)}$ and $\chi_{p,s}^{(n)}$ still depend in an involved way on the coupling and dressing fields as evident from Eq.~(\ref{equ8}) and the four-channel scattering processes they describe (in particular the cross ones) stem from nonlinear wave-mixing effects.

\begin{figure}
\begin{centering}
\includegraphics[width=0.48\textwidth]{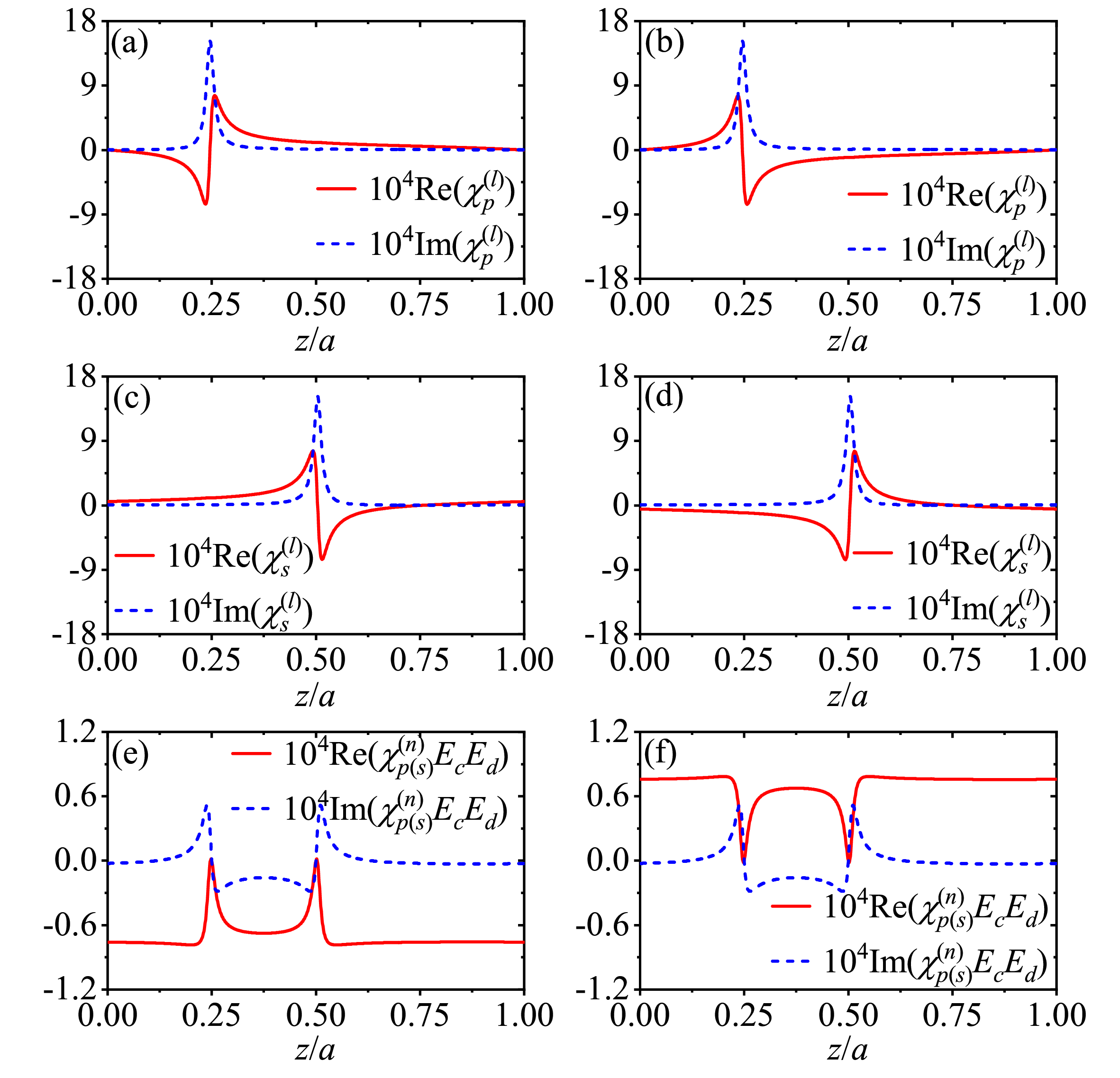}
\caption{Susceptibilities $\chi_{p}^{(l)}$ (a, b), $\chi_{s}^{(l)}$ (c, d), and $\chi_{p}^{(n)}=\chi_{s}^{(n)}$ (e, f) vs atomic position $z$ attained with the same parameters as in Fig.~\ref{fig4} except $\Delta_{p}=2\pi\times43$ MHz in (a, c, e) on the left while $\Delta_{p}=-2\pi\times43$ MHz in (b, d, f) on the right.}\label{fig6}
\end{centering}
\end{figure}

In the case of $\Delta_{c}=\Delta_{d}$ and $\phi=0$, it is easy to learn from Eq.~(\ref{equ8}) that $A(z)=C(z)$ and $B(z)=D(z)$, thereby we must arrive at $\chi_{p}^{(l)}=\chi_{s}^{(l)}$ and $\chi_{p}^{(n)}=\chi_{s}^{(n)}$, which is why all reflections and transmissions are symmetric, \textit{i.e.} invariant upon an exchange of the probe and signal fields. It is worth noting that $\chi_{p}^{(l)}=\chi_{s}^{(l)}$ and $\chi_{p}^{(n)}=\chi_{s}^{(n)}$ are also Hermitian with $\mathrm{Im}[\chi_{p,s}^{(l,n)}]$ and $\mathrm{Re}[\chi_{p,s}^{(l,n)}]$ being spatially modulated in phase, hence all reflections and transmissions are reciprocal, \textit{i.e.} invariant upon an exchange of the input and output ports. Nonreciprocal transmissions and reflections can only be attained with non-Hermitian susceptibilities in the case of $\phi\ne k\pi$, and will become most pronounced when $\mathrm{Im}[\chi_{p,s}^{(l,n)}]$ and $\mathrm{Re}[\chi_{p,s}^{(l,n)}]$ exhibit the largest spatial phase mismatch for $\phi=(k\pm1/4)\pi$, of course depending also on whether $\Delta_{c}=\Delta_{d}$.

In Fig.~\ref{fig6}, we plot the linear and nonlinear susceptibilities against position $z$ in a single period of our atomic sample at $\Delta_{p}/2\pi=\pm43$ MHz with the same parameters as in Fig.~\ref{fig4}. It is easy to observe that linear susceptibilities $\chi_{p}^{(l)}(z)$ and $\chi_{s}^{(l)}(z)$ exhibit exact $\mathcal{PT}$ antisymmetric resonances~\cite{PRL.113.123004} because their imaginary (real) parts are even (odd) functions with respect to respective resonance centers. It is also clear that $\chi_{p}^{(l)}(z)$ and $\chi_{s}^{(l)}(z)$ are similar in profile with identical imaginary parts but opposite real parts though staggered by a $1/4$ period in space. This is why nonreciprocal and even unidirectional direct reflections have been observed for the probe and signal fields due to $|r_{pp}^{+-}|\ne|r_{pp}^{-+}|$ and $|r_{ss}^{+-}|\ne|r_{ss}^{-+}|$, and why they are reversed in terms of input and output directions due to $|r_{pp}^{+-}|=|r_{ss}^{-+}|$ and $|r_{pp}^{-+}|=|r_{ss}^{+-}|$. On the other hand, nonlinear susceptibilities $\chi_{p}^{(n)}(z)=\chi_{s}^{(n)}(z)$ exhibit two spatially staggered resonances in accordance to those of $\chi_{p}^{(l)}(z)$ and $\chi_{s}^{(l)}(z)$, respectively, but are just partially $\mathcal{PT}$ antisymmetric with their real (imaginary) part deviating from an odd (even) function with respect to the center of each resonance. It is the interplay between exact $\mathcal{PT}$ antisymmetric linear susceptibilities and partially $\mathcal{PT}$ antisymmetric nonlinear susceptibilities that results in nonreciprocal cross transmissions. For attaining nonreciprocal cross reflections, we should further destroy the similarity between two linear susceptibilities by introducing asymmetric driving fields with $\Delta_{c}\ne\Delta_{d}$.

Above discussions according to numerical evidences could be substantiated by straightforward, albeit laborious, analytical calculations based on an expansion and a truncation of the transfer matrix $\hat{M}$ as detailed in Appendix~\ref{appC}. These calculations first tell that the truncated transfer matrix and hence all transmission and reflection amplitudes are clearly expressed in terms of the $0$th-order and $\pm1$st-order Fourier components of four linear $\chi_{p,s}^{(l)}(z)$ and nonlinear $\chi_{p,s}^{(n)}(z)$ susceptibilities. Non-Hermitian scattering then should occur when the $1$st-order and $-1$st-order Fourier components become different for each susceptibility in the case of $\phi\ne k\pi$, thereby leading to nonreciprocal direct reflections, cross reflections, and cross transmissions in general, while direct transmissions are intrinsically reciprocal. One exception is that cross reflections may happen to be reciprocal when the coupling and dressing fields are applied in a symmetric way so that the $0$th-order and $\pm1$st-order Fourier components also exhibit a certain exchange symmetry. Finally, we note that it is enough to capture most features of direct reflections, cross reflections, and direct transmissions with the first-order truncation while nonreciprocal cross transmissions won't appear until the second-order truncation.

\section{\label{sec5}Conclusions}

Optical nonreciprocity is a long-standing phenomenon of fundamental interest, yet today largely driven by applications of nonreciprocal devices in areas such as, \textit{e.g.} signal processing and quantum networks. Within this context, achieving the full control over nonreciprocity has been a highly desirable and remarkable feature for any such devices. We here harness third-order nonlinearities and induced Bragg scattering in a familiar multi-level (double-$\Lambda$) configuration to attain nonreciprocal ``direct" reflections as well as nonreciprocal ``cross" reflections and transmissions, where ``direct (cross)" refers to a beam scattering off a \textit{single-input} port onto another one of the same (different) frequency. In a specific atomic medium, which we suitably engineer to implement our proposal, it is possible to adjust parameters of the medium's driving scheme to switch between these nonreciprocal scattering channels. By the same means, we have also encompassed both Hermitian and non-Hermitian behaviors, making our results a valuable contribution to the understanding of non-Hermitian optical scattering. Finally, the present atomic archetype may possibly be operated in a regime where more than one input-port is used (unlike in this work). Our results would establish then the ground for investigating multi-color optical nonreciprocity within a single device, further enhancing their significance.

\section*{ACKNOWLEDGMENTS}
This work is supported by the National Natural Science Foundation of China (Grants No.~62375047 and No.~12074061), the Italian PNRR MUR (No.~PE0000023-NQSTI), I-PHOQS (Photonics and Quantum Sciences, PdGP/GePro 2024-2026), and the Fund for International Activities of the University of Brescia.

\appendix
\section{\label{appA} Equations for TW coupling and dressing fields}
It is known that the propagation dynamics of a light field with amplitude $E$ inside a medium with polarization $P$ is governed by the Maxwell equation
\begin{equation}\label{A1}
\frac{\partial^{2}E}{\partial z^{2}}-\frac{1}{c^2}\frac{\partial^{2}E}{\partial t^{2}}=\mu_{0}\frac{\partial^{2}P}{\partial t^{2}},
\end{equation}
when $E$ and $P$ oscillate in time (space) with roughly the same frequency (wavenumber) $\omega$ ($k=\omega/c$). Now we focus on the double-$\Lambda$ atomic system in Fig.~\ref{fig1}(b) and consider the simpler case where both coupling and dressing beams are in the TW pattern and travel in the forward direction with Rabi frequencies $\Omega_{c}e^{ik_{c}z}$ and $\Omega_{d}e^{ik_{d}z}$. Under the slowly-varying envelope approximation, it is viable to reduce the above Maxwell equation into
\begin{eqnarray}\label{A2}
\frac{\partial E_{p}}{\partial z}e^{ik_{p}z} &=& \eta_{p}[P_{31}^{(l)}e^{ik_{p}z}+P_{31}^{(n)}e^{i(k_{p}-\Delta k)z}],\nonumber \\
\frac{\partial E_{s}}{\partial z}e^{ik_{s}z} &=& \eta_{s}[P_{41}^{(l)}e^{ik_{s}z}+P_{41}^{(n)}e^{i(k_{s}+\Delta k)z}],
\end{eqnarray}
with $\eta_{p}=i\omega_{p}/2\varepsilon_{0}c$ and $\eta_{s}=i\omega_{s}/2\varepsilon_{0}c$ in the steady state. Here, $E$ is replaced by $E_{p}e^{ik_{p}z}$ and $E_{s}e^{ik_{s}z}$ for a probe and a signal fields both travelling in the forward direction, and accordingly $P$ by $P_{31}^{(l)}e^{ik_{p}z}+P_{31}^{(n)}e^{i(k_{p}-\Delta k)z}$ and $P_{41}^{(l)}e^{ik_{s}z}+P_{41}^{(n)}e^{i(k_{s}+\Delta k)z}$ with a wavenumber difference $\Delta k=k_{p}-k_{c}+k_{d}-k_{s}$, which would result in a phase mismatch during light propagation. Relevant linear and nonlinear polarizations are given by
\begin{eqnarray}\label{A3}
P_{31}^{(l)}&=& Nd_{13}\rho_{31}^{(l)}=\alpha_{13}AE_{p}, \nonumber \\
P_{31}^{(n)}&=& Nd_{13}\rho_{31}^{(n)}=\alpha_{13}BE_{s}, \nonumber \\
P_{41}^{(l)}&=& Nd_{14}\rho_{41}^{(l)}=\alpha_{14}CE_{s}, \nonumber \\
P_{41}^{(n)}&=& Nd_{14}\rho_{41}^{(n)}=\alpha_{14}DE_{p},
\end{eqnarray}
with $A$, $B$, $C$, and $D$ defined as in Eq.~(\ref{equ8}) but becoming space-invariant here. Moreover, $\alpha_{13}=Nd_{13}^{2}/2\hbar$ and $\alpha_{14}=Nd_{14}^{2}/2\hbar$ are introduced for convenience.

In the ideal case of perfect phase matching ($\Delta k=0$), substituting Eq.~(\ref{A3}) into Eq.~(\ref{A2}), we further attain the following two-mode coupled equations
\begin{equation}\label{A4}
L\frac{\partial}{\partial z}\left({\begin{array}{*{20}{c}}
E_{p}\\
E_{s}
\end{array}}\right) = \hat{X}_{T}\left({\begin{array}{*{20}{c}}
E_{p}\\
E_{s}
\end{array}}\right) = \left({\begin{array}{*{20}{c}}
\mathcal{A} & \mathcal{B}\\
\mathcal{D} & \mathcal{C}
\end{array}}\right) \left({\begin{array}{*{20}{c}}
E_{p}\\
E_{s}
\end{array}}\right),
\end{equation}
with $\mathcal{A}/A=\mathcal{B}/B=\eta_{p}\alpha_{13}L$ and $\mathcal{C}/C=\mathcal{D}/D=\eta_{s}\alpha_{14}L$. A formal integration of this equation yields
\begin{equation}\label{A5}
\left({\begin{array}{*{20}{c}}
{E_{p}}(L)\\
{E_{s}}(L)
\end{array}} \right) = \hat{M}^{T}\left({\begin{array}{*{20}{c}}
{E_{p}}(0)\\
{E_{s}}(0)
\end{array}} \right),
\end{equation}
where $\hat{M}^{T}=e^{\hat{X}_{T}}$ is a $2\times2$ transfer matrix composed of four elements $M^{T}_{11}$, $M^{T}_{12}$, $M^{T}_{21}$, and $M^{T}_{22}$.

\begin{figure}
\centering
\includegraphics[width=0.48\textwidth]{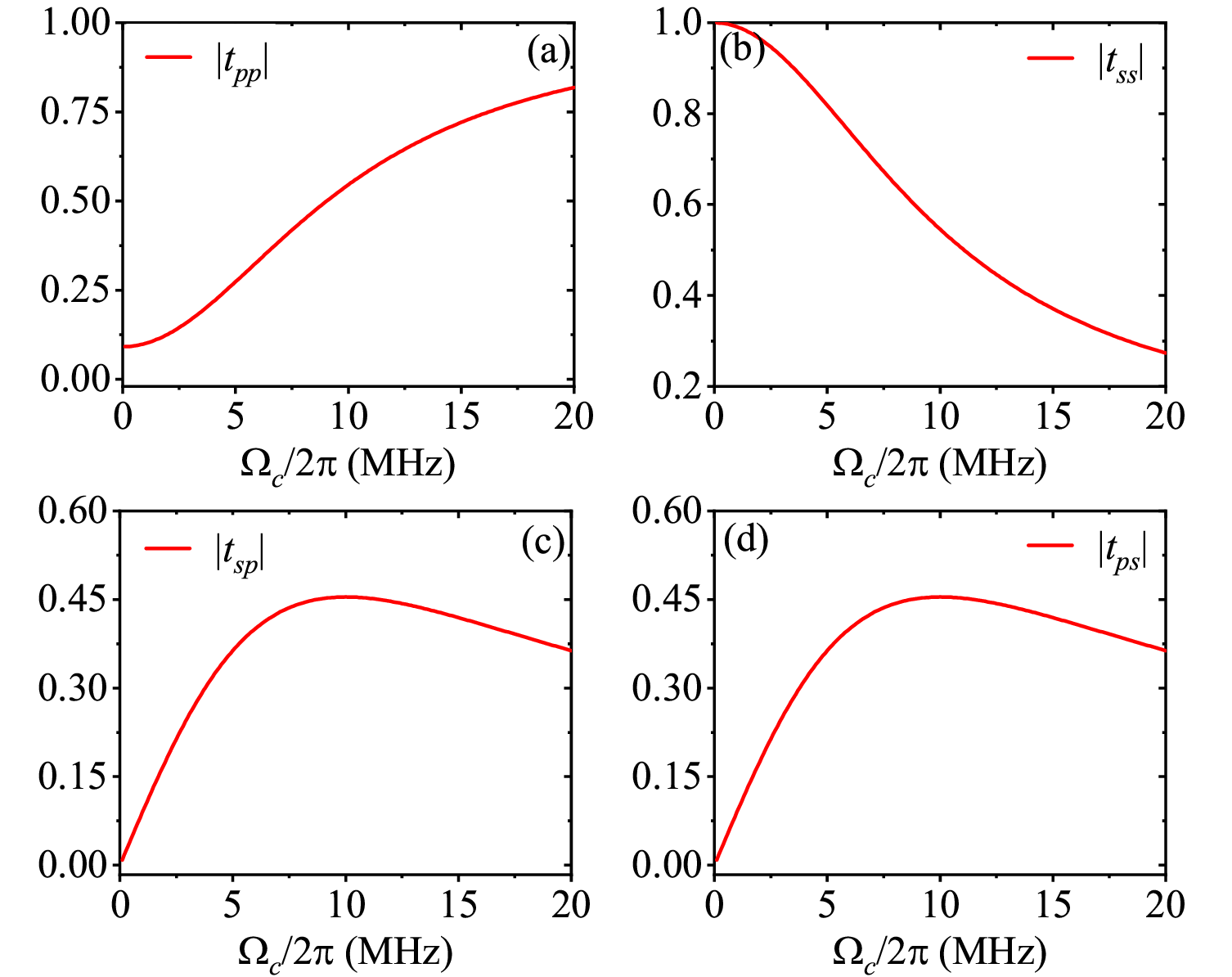}\\
\caption{Moduli of transmission amplitudes vs Rabi frequency $\Omega_{c}$ for a finite atomic sample driven by the TW coupling and dressing fields. Relevant parameters are the same as in Fig.~\ref{fig2} except $\Delta_{p}=\Delta_{s}=0$ and $\Omega_{d}=2\pi\times10$ MHz.}\label{figA1}
\end{figure}

Based on Eq.~(\ref{A5}), it is viable to ultimately calculate the direct and cross transmission amplitudes
\begin{eqnarray}\label{A6}
t_{pp} &=& \frac{E_{p}(L)}{E_{p}(0)}=M_{11}^{T},\,\ t_{ss}=\frac{E_{s}(L)}{E_{s}(0)}=M_{22}^{T}, \nonumber\\
t_{ps} &=& \frac{E_{s}(L)}{E_{p}(0)}=M_{21}^{T},\,\ t_{sp}=\frac{E_{p}(L)}{E_{s}(0)}=M_{12}^{T},
\end{eqnarray}
which are very simple because forward probe and signal fields won't be scattered into backward ones. Considering a backward incidence with $E_{p}e^{-ik_{p}z}$ and $E_{s}e^{-ik_{s}z}$, we can attain the same transmission amplitudes as a result of the Lorentz reciprocity theorem~\cite{LRT}. In Fig.~\ref{figA1}, the moduli of transmission amplitudes $|t_{pp}|$, $|t_{ss}|$, $|t_{ps}|$, and $|t_{sp}|$ are plotted against the coupling Rabi frequency $\Omega_{c}$. It is easy to see that direct transmission $|t_{pp}|$ of the probe field gradually increases from $0.1$ (resonant absorption) while direct transmission $|t_{ss}|$ of the signal field gradually decreases from $1.0$ (EIT effect) when $\Omega_{c}$ is turned on and rises continuously. We also find that cross transmissions $|t_{ps}|$ and $|t_{sp}|$ answering for nonlinear conversions are always equal to each other due to $\Delta_{c}=\Delta_{d}$, both vanish in the case of $\Omega_{c}=0$, and reach a maximum for balanced coupling and dressing fields with $\Omega_{c}=\Omega_{d}$. These findings are consistent with those in ref.~\cite{PRA.89.023839}.

\section{\label{appB} Equations for SW coupling and dressing fields}
In the more complicated case where the coupling and dressing beams are in the SW pattern with electric fields $E_{c}(z)$ and $E_{d}(z)$ described by Eq.~(\ref{equ6}), the Maxwell equations reduced in the slowly-varying envelope approximation and in the steady state turn out to be
\begin{eqnarray}\label{B1}
\frac{\partial E_{p}^{+}}{\partial z}e^{ik_{p}z} - \frac{\partial E_{p}^{-}}{\partial z}e^{-ik_{p}z} &=& \eta_{p}P_{31}(z),\nonumber \\
\frac{\partial E_{s}^{+}}{\partial z}e^{ik_{s}z} - \frac{\partial E_{s}^{-}}{\partial z}e^{-ik_{s}z} &=& \eta_{s}P_{41}(z).
\end{eqnarray}
Here both probe and signal fields are assumed to contain a forward and a backward components because they fulfill the phase-matching requirement $\Delta k=0$ with different Fourier components of the periodic polarizations $P_{31}(z)$ and $P_{41}(z)$ in Eq.~(\ref{equ10}). Assuming $k_{c}=k_{d}=k_{0}$ as in the main text, we expand the space-dependent terms $\mathcal{A}(z)=\eta_{p}\alpha_{13}LA(z)\propto\chi_{p}^{(l)}(z)$, $\mathcal{B}(z)=\eta_{p}\alpha_{13}LB(z)\propto\chi_{p}^{(n)}(z)$, $\mathcal{C}(z)=\eta_{s}\alpha_{14}LC(z)\propto\chi_{s}^{(l)}(z)$, and $\mathcal{D}(z)=\eta_{s}\alpha_{14}LD(z)\propto\chi_{s}^{(n)}(z)$ into the Fourier series
\begin{eqnarray}\label{B2}
\mathcal{A}(z) &=& \mathcal{A}_{0}+\mathcal{A}_{1+}e^{i2k_{0}z}+\mathcal{A}_{1-}e^{-i2k_{0}z}+..., \nonumber \\
\mathcal{B}(z) &=& \mathcal{B}_{0}+\mathcal{B}_{1+}e^{i2k_{0}z}+\mathcal{B}_{1-}e^{-i2k_{0}z}+..., \nonumber \\
\mathcal{C}(z) &=& \mathcal{C}_{0}+\mathcal{C}_{1+}e^{i2k_{0}z}+\mathcal{C}_{1-}e^{-i2k_{0}z}+..., \nonumber\\
\mathcal{D}(z) &=& \mathcal{D}_{0}+\mathcal{D}_{1+}e^{i2k_{0}z}+\mathcal{D}_{1-}e^{-i2k_{0}z}+...,
\end{eqnarray}
where the zeroth-order $\Upsilon_{0}$ and the first-order $\Upsilon_{1\pm}$ components with $\Upsilon\in\{\mathcal{A},\mathcal{B},\mathcal{C},\mathcal{D}\}$ can be calculated through $\Upsilon_{0}=\frac{1}{a} \int_{0}^{a}\Upsilon(z) dz$ and $\Upsilon_{1\pm}=\frac{1}{a}\int_{0}^{a} \Upsilon(z)e^{\pm i2k_{0}z}dz$ with $a=\pi/k_{0}$ being the common period of SW coupling and dressing fields. Taking the space-dependent polarizations in Eq.~(\ref{equ10}) and coefficients in Eq.~(\ref{B2}) into Eq.~(\ref{B1}), we can derive the four-mode coupled equations
\begin{equation}\label{B3}
L\frac{\partial}{\partial z} \left( {\begin{array}{*{20}{c}}
{E_{p}^{+}}\\
{E_{p}^{-}}\\
{E_{s}^{+}}\\
{E_{s}^{-}}
\end{array}} \right) = \hat{X} \left( {\begin{array}{*{20}{c}}
{E_{p}^{+}}\\
{E_{p}^{-}}\\
{E_{s}^{+}}\\
{E_{s}^{-}}
\end{array}} \right),
\end{equation}
where the coefficient matrix $\hat{X}$ is given by
\begin{equation}
\hat{X} = \left( {\begin{array}{*{20}{c}}
\mathcal{A}_{0} & \mathcal{A}_{1-} & \mathcal{B}_{0} & \mathcal{B}_{1-}\\
-\mathcal{A}_{1+} & -\mathcal{A}_{0} & -\mathcal{B}_{1+} & -\mathcal{B}_{0}\\
\mathcal{D}_{0} & \mathcal{D}_{1-} & \mathcal{C}_{0} & \mathcal{C}_{1-}\\
-\mathcal{D}_{1+} & -\mathcal{D}_{0} & -\mathcal{C}_{1+} & -\mathcal{C}_{0}
\end{array}} \right).\nonumber
\end{equation}

A formal integration of Eq.~(\ref{B3}) then leads to Eq.~(\ref{equ5}) in the main text where a $4\times4$ transfer matrix $\hat{M}$ relates the probe and signal fields at $z=L$ to other two at $z=0$. Transfer matrix elements $M_{ij}$ and relevant boundary conditions~\cite{bc} allow us to ultimately calculate
\begin{widetext}
\begin{eqnarray}\label{B4}
r_{pp}^{+-} &=& \frac{M_{24}M_{41} - M_{21}M_{44}} {M_{22}M_{44} - M_{24}M_{42}}, \,\
r_{pp}^{-+} = \frac{M_{12}M_{44} - M_{14}M_{42}} {M_{22}M_{44} - M_{24}M_{42}}, \nonumber \\
r_{ss}^{+-} &=& \frac{M_{23}M_{42} - M_{22}M_{43}} {M_{22}M_{44} - M_{24}M_{42}}, \,\
r_{ss}^{-+} = \frac{M_{22}M_{34} - M_{24}M_{32}} {M_{22}M_{44} - M_{24}M_{42}},
\end{eqnarray}
referring to the direct reflection amplitudes of probe and signal fields coming from $\pm z$ directions;
\begin{eqnarray}\label{B5}
r_{ps}^{+-} &=& \frac{M_{21}M_{42} - M_{22}M_{41}} {M_{22}M_{44} - M_{24}M_{42}}, \,\
r_{ps}^{-+} = \frac{M_{32}M_{44} - M_{34}M_{42}} {M_{22}M_{44} - M_{24}M_{42}},\nonumber\\
r_{sp}^{+-} &=& \frac{M_{24}M_{43} - M_{23}M_{44}} {M_{22}M_{44} - M_{24}M_{42}}, \,\
r_{sp}^{-+} = \frac{M_{14}M_{22} - M_{12}M_{24}} {M_{22}M_{44} - M_{24}M_{42}},
\end{eqnarray}
referring to the cross reflection amplitudes of probe and signal fields coming from $\pm z$ directions;
\begin{eqnarray}\label{B6}
t_{pp}^{++} &=& M_{11} + \frac{M_{12}(M_{24}M_{41} - M_{21}M_{44}) + M_{14}(M_{21}M_{42} - M_{22}M_{41})} {M_{22}M_{44} - M_{24}M_{42}}, \nonumber \\
t_{ss}^{++} &=& M_{33} + \frac{M_{32}(M_{24}M_{43} - M_{23}M_{44}) + M_{34}(M_{23}M_{42} - M_{22}M_{43})} {M_{22}M_{44} - M_{24}M_{42}}, \nonumber \\
t_{pp}^{--} &=& \frac{M_{44}} {M_{22}M_{44} - M_{24}M_{42}},\,\
t_{ss}^{--} = \frac{M_{22}} {M_{22}M_{44} - M_{24}M_{42}},
\end{eqnarray}
referring to the direct transmission amplitudes of probe and signal fields coming from $\pm z$ directions;
\begin{eqnarray}\label{B7}
t_{ps}^{++} &=& M_{31} + \frac{M_{21}(M_{42}M_{34} - M_{44}M_{32}) + M_{41}(M_{24}M_{32} - M_{22}M_{34})} {M_{22}M_{44} - M_{24}M_{42}}, \nonumber \\
t_{sp}^{++} &=& M_{13} + \frac{M_{12}(M_{24}M_{43} - M_{44}M_{23}) + M_{14}(M_{42}M_{23} - M_{22}M_{43})} {M_{22}M_{44} - M_{24}M_{42}}, \nonumber \\
t_{ps}^{--} &=& - \frac{M_{42}} {M_{22}M_{44} - M_{24}M_{42}},\,\
t_{sp}^{--} = - \frac{M_{24}} {M_{22}M_{44} - M_{24}M_{42}},
\end{eqnarray}
referring to the cross transmission amplitudes of probe and signal fields coming from $\pm z$ directions.
\end{widetext}

\begin{figure*}
\begin{centering}
\includegraphics[width=0.96\textwidth]{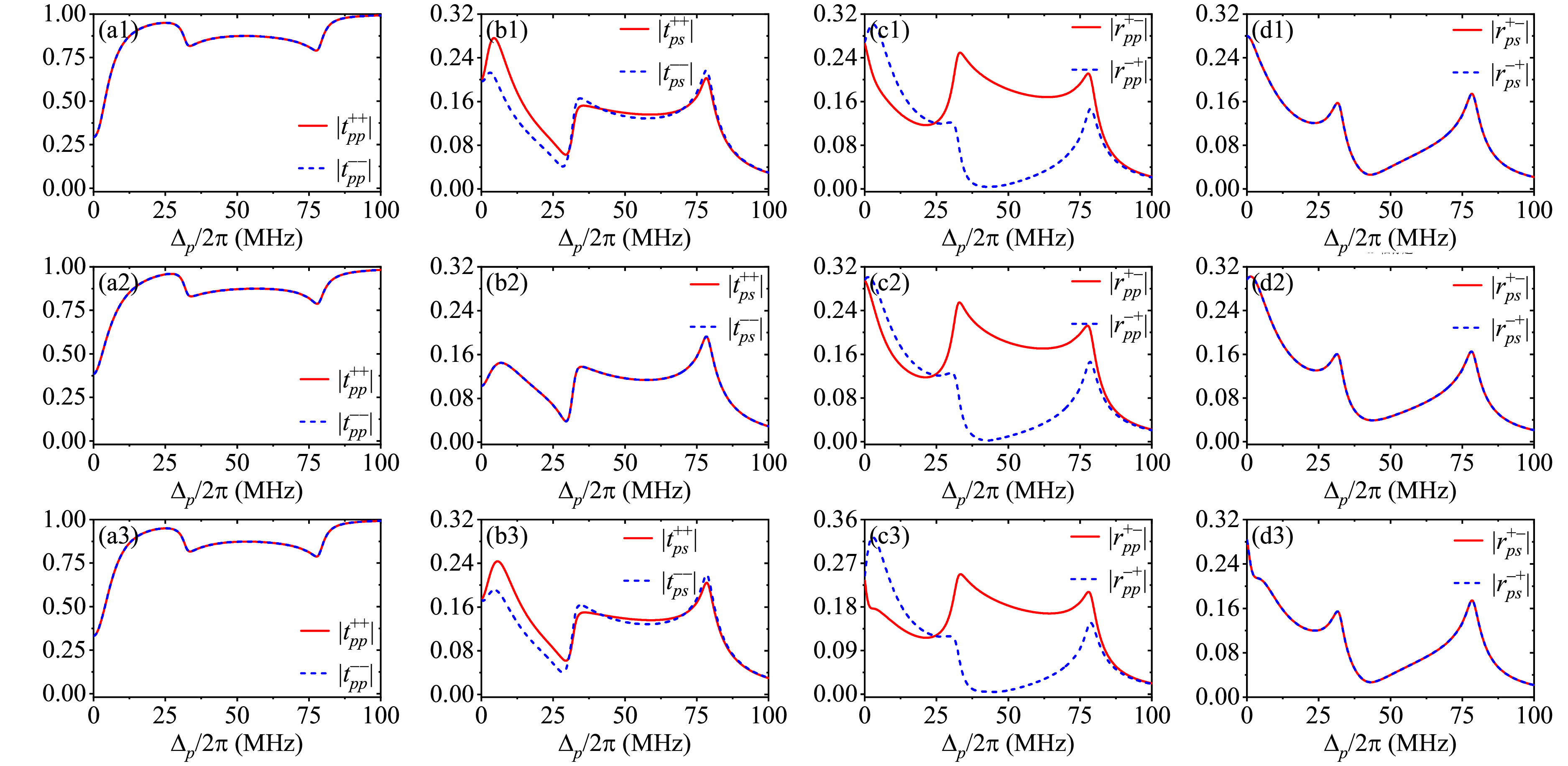}\\
\caption{Moduli of direct (a1-a3) and cross (b1-b3) transmission amplitudes as well as direct (c1-c3) and cross (d1-d3) reflection amplitudes vs probe detuning $\Delta_{p}$ attained with the same parameters as in Fig.~\ref{fig4}. Lines in (a1-d1) are exact results while those in (a2-d2) and (a3-d3) refer to transfer matrices truncated at first-order and second-order Taylor expansions, respectively. }\label{figA2}
\end{centering}
\end{figure*}

\section{\label{appC} Expansion and truncation of transfer matrix}
In order to better understand the nonreciprocal scattering behaviors observed in Fig.~\ref{fig3}--Fig.~\ref{fig5}, below we try to derive approximate analytical expressions for all sixteen reflection and transmission amplitudes. This seemingly formidable task can be accomplished by expanding the transfer matrix $\hat{M}$ into a Taylor series
\begin{equation}\label{C1}
\hat{M} = e^{\hat{X}} = \hat{I} + \hat{X} + \hat{X}^{2}/2! + \hat{X}^{3}/3! +...,
\end{equation}
and making truncations to the first and second orders of $\hat{X}$. The accuracy of such truncations for appropriate parameters is verified in Fig.~\ref{figA2} by comparing exact results to both first-order and second-order approximate results. It is clear that first-order approximations on direct transmissions, direct reflections, and cross reflections have already been accurate enough, whereas second-order approximations should be adopted to attain similarly accurate cross transmissions. That means, $\hat{X}^{3}/3!$ and other higher-order terms don't contribute essentially to direct and cross reflections as well as direct and cross transmissions. Taking the first-order truncation $\hat{M}\simeq\hat{I}+\hat{X}$, we derive the approximate reflection amplitudes
\begin{eqnarray}\label{C2}
\tilde{r}_{pp}^{+-} &=& \frac{\mathcal{B}_{0}\mathcal{D}_{1+} + \mathcal{A}_{1+}(1-\mathcal{C}_{0})}
{(1-\mathcal{A}_{0})(1-\mathcal{C}_{0}) - \mathcal{B}_{0}\mathcal{D}_{0}}, \nonumber \\
\tilde{r}_{pp}^{-+} &=& \frac{\mathcal{D}_{0}\mathcal{B}_{1-} + \mathcal{A}_{1-}(1-\mathcal{C}_{0})}
{(1-\mathcal{A}_{0})(1-\mathcal{C}_{0}) - \mathcal{B}_{0}\mathcal{D}_{0}}, \nonumber  \\
\tilde{r}_{ps}^{+-} &=& \frac{\mathcal{D}_{0}\mathcal{A}_{1+} + \mathcal{D}_{1+}(1-\mathcal{A}_{0})}
{(1-\mathcal{A}_{0})(1-\mathcal{C}_{0}) - \mathcal{B}_{0}\mathcal{D}_{0}}, \nonumber \\
\tilde{r}_{ps}^{-+} &=& \frac{\mathcal{D}_{0}\mathcal{C}_{1-} + \mathcal{D}_{1-}(1-\mathcal{C}_{0})}
{(1-\mathcal{A}_{0})(1-\mathcal{C}_{0}) - \mathcal{B}_{0}\mathcal{D}_{0}}.
\end{eqnarray}
With the second-order truncation $\hat{M}\simeq\hat{I}+\hat{X}+\hat{X}^{2}/2!$, we get the approximate transmission amplitudes
\begin{eqnarray}\label{C3}
\tilde{t}_{pp}^{++} = \tilde{t}_{pp}^{--} &=& \frac{1-\mathcal{C}_{0}^{\prime}}{(1-\mathcal{A}_{0}^{\prime})(1-\mathcal{C}_{0}^{\prime})
 -\mathcal{B}_{0}^{\prime}\mathcal{D}_{0}^{+}}, \nonumber \\
\tilde{t}_{ps}^{++} &=& \frac{\mathcal{D}_{0}^{+}}{(1-\mathcal{A}_{0}^{\prime})(1-\mathcal{C}_{0}^{\prime})
 -\mathcal{B}_{0}^{\prime}\mathcal{D}_{0}^{+}}, \nonumber \\
\tilde{t}_{ps}^{--} &=& \frac{\mathcal{D}_{0}^{-}}{(1-\mathcal{A}_{0}^{\prime})(1-\mathcal{C}_{0}^{\prime})
 -\mathcal{B}_{0}^{\prime}\mathcal{D}_{0}^{+}},
\end{eqnarray}
where five corrected parameters are defined as
\begin{eqnarray}\label{C4}
\mathcal{A}_{0}^{\prime} &=&\mathcal{A}_{0}-  \frac{\mathcal{A}_{0}^{2}+\mathcal{B}_{0}\mathcal{D}_{0}-\mathcal{A}_{1+}\mathcal{A}_{1-}-\mathcal{B}_{1+}\mathcal{D}_{1-}}{2}, \nonumber \\
\mathcal{B}_{0}^{\prime} &=&\mathcal{B}_{0}-  \frac{\mathcal{A}_{0}\mathcal{B}_{0}+\mathcal{B}_{0}\mathcal{C}_{0}-\mathcal{A}_{1+}\mathcal{B}_{1-}-\mathcal{B}_{1+}\mathcal{C}_{1-}}{2}, \nonumber \\
\mathcal{C}_{0}^{\prime} &=&\mathcal{C}_{0}-  \frac{\mathcal{C}_{0}^{2}+\mathcal{B}_{0}\mathcal{D}_{0}-\mathcal{C}_{1+}\mathcal{C}_{1-}-\mathcal{D}_{1+}\mathcal{B}_{1-}}{2}, \\
\mathcal{D}_{0}^{+} &=&\mathcal{D}_{0}- \frac{\mathcal{A}_{0}\mathcal{D}_{0}+\mathcal{C}_{0}\mathcal{D}_{0}-\mathcal{C}_{1+}\mathcal{D}_{1-}-\mathcal{A}_{1-}\mathcal{D}_{1+}}{2}, \nonumber \\
\mathcal{D}_{0}^{-} &=&\mathcal{D}_{0}- \frac{\mathcal{A}_{0}\mathcal{D}_{0}+\mathcal{C}_{0}\mathcal{D}_{0}-\mathcal{A}_{1+}\mathcal{D}_{1-}-\mathcal{C}_{1-}\mathcal{D}_{1+}}{2}. \nonumber
\end{eqnarray}

\begin{figure*}
\begin{centering}
\includegraphics[width=0.96\textwidth]{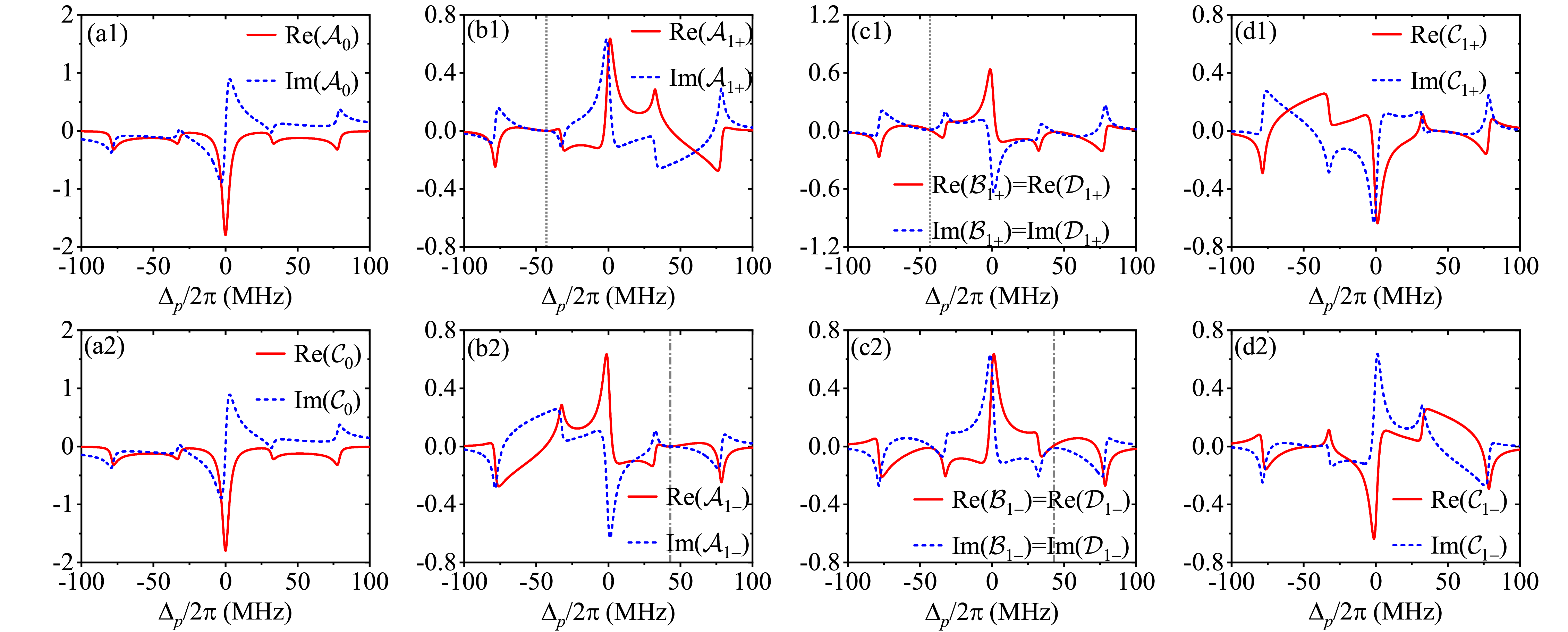}
\caption{Elements $\mathcal{A}_{0}$ (a1), $\mathcal{A}_{1+}$ (b1), $\mathcal{B}_{1+}=\mathcal{D}_{1+}$ (c1), and $\mathcal{C}_{1+}$ (d1) as well as $\mathcal{C}_{0}$ (a2), $\mathcal{A}_{1-}$ (b2), $\mathcal{B}_{1-}=\mathcal{D}_{1-}$ (c2), and $\mathcal{C}_{1-}$ (d2) in the coefficient matrix $\hat{X}$ vs probe detuning $\Delta_{p}$ attained with the same parameters as in Fig.~\ref{fig4}.}\label{figA3}
\end{centering}
\end{figure*}

For simplicity, we restrict our following discussions to the case of $\Delta_{c}=\Delta_{d}$ and $\gamma_{31}=\gamma_{41}=\gamma$. Further taking $\phi=0$, it is easy to find from Eq.~(\ref{equ8}) that $A(z)=C(z)$ and $B(z)=D(z)$, thereby we should arrive at $\mathcal{A}_{0}=\mathcal{C}_{0}$, $\mathcal{B}_{0}=\mathcal{D}_{0}$, $\mathcal{A}_{1\pm}=\mathcal{C}_{1\pm}$, and $\mathcal{B}_{1\pm}=\mathcal{D}_{1\pm}$. Accordingly, all reflections and transmissions must be reciprocal as shown in Fig.~\ref{fig2}, where the plotted curves are also symmetric with respect to $\Delta_{p}=0$. Taking $\phi=\pi/4$ as in Fig.~\ref{fig3} and Fig.~\ref{fig4}, we plot relevant elements of the coefficient matrix $\hat{X}$ against probe detuning $\Delta_{p}$ in Fig.~\ref{figA3}, which shows that $\mathcal{A}_{0}=\mathcal{C}_{0}$, $\mathcal{A}_{1-}=i\mathcal{C}_{1+}$, $\mathcal{B}_{1-}=i\mathcal{B}_{1+}$, and $\mathcal{C}_{1-}=i\mathcal{A}_{1+}$. Further considering $\mathcal{D}_{0}\equiv\mathcal{B}_{0}$ and $\mathcal{D}_{1\pm}\equiv\mathcal{B}_{1\pm}$ in our model, we learn from Eq.~(\ref{C2}) that $|\tilde{r}_{ps}^{+-}|=|\tilde{r}_{ps}^{-+}|$ due to $\mathcal{A}_{1+}/\mathcal{C}_{1-}=\mathcal{D}_{1+}/\mathcal{D}_{1-}$ and $\mathcal{A}_{0}=\mathcal{C}_{0}$ while $|\tilde{r}_{pp}^{+-}|\ne|\tilde{r}_{pp}^{-+}|$ due to $\mathcal{D}_{1+}/\mathcal{D}_{1-}\ne\mathcal{A}_{1+}/\mathcal{A}_{1-}$ and $\mathcal{D}_{0}=\mathcal{B}_{0}$. We also learn from Eq.~(\ref{C3}) that $|\tilde{t}_{pp}^{+-}|=|\tilde{t}_{pp}^{-+}|$ always holds true while $|\tilde{t}_{ps}^{+-}|\ne|\tilde{t}_{ps}^{-+}|$ is attained because of $\mathcal{A}_{1\pm}\ne\mathcal{C}_{1\pm}$ with $\mathcal{A}_{1-}=i\mathcal{C}_{1+}$ and $\mathcal{C}_{1-}=i\mathcal{A}_{1+}$ as shown by Fig.~\ref{figA3}. Note, however, that we have $|\tilde{t}_{ps}^{+-}|=|\tilde{t}_{ps}^{-+}|$ when both $\mathcal{D}_{0}^{+}$ and $\mathcal{D}_{0}^{-}$ are replaced by $\mathcal{D}_{0}$ for the first-order truncation of transfer matrix $\hat{M}$. That means, it is the second-order scattering effect that results in $|\tilde{t}_{ps}^{+-}|-|\tilde{t}_{ps}^{-+}|\ne0$ so that it is less evident than $|\tilde{r}_{pp}^{+-}|-|\tilde{r}_{pp}^{-+}|\ne0$. Above discussions well explain how linear $\{\mathcal{A}_{0},\mathcal{A}_{1\pm},\mathcal{C}_{0},\mathcal{C}_{1\pm}\}$ and nonlinear  $\{\mathcal{B}_{0},\mathcal{B}_{1\pm},\mathcal{D}_{0},\mathcal{D}_{1\pm}\}$ contributions jointly cause nonreciprocal direct reflection and cross transmission while reciprocal cross reflection and direct transmission.

Above numerical results can be further verified by the following analytical expressions. In the case of a symmetric driving considered here, we can define $\digamma=\eta_{p}\alpha_{13}L=\eta_{s}\alpha_{14}L$ and attain from Eq.~(\ref{equ8}) and Eq.~(\ref{B2})
\begin{eqnarray}\label{C5}
\mathcal{A}(z)&=&\frac{i\digamma}{g_{31}}\frac{[1-\sin(2kz)]+\beta}{[2+\cos(2kz)-\sin(2kz)]+\beta}, \nonumber \\
\mathcal{B}(z)&=&\frac{i\digamma}{g_{31}}\frac{-[1+\cos(2kz)-\sin(2kz)]/\sqrt{2}}{[2+\cos(2kz)-\sin(2kz)]+\beta}, \nonumber \\
\mathcal{C}(z)&=&\frac{i\digamma}{g_{31}}\frac{[1+\cos(2kz)]+\beta}{[2+\cos(2kz)-\sin(2kz)]+\beta},
\end{eqnarray}
with $\mathcal{D}(z)=\mathcal{B}(z)$ and $\beta=g_{21}g_{31}/G^{2}$. Then, it is viable to further attain their Fourier components
\begin{eqnarray}\label{C6}
\mathcal{A}_{0}&=&\mathcal{C}_{0}=\frac{-[(1+i)(1+\beta)+iz_{-}]\digamma}{(z_{+}-z_{-})g_{31}}, \nonumber \\
\mathcal{B}_{0}&=&\frac{\sqrt{2}[(1+i)-2iz_{-}]\digamma+}{2(z_{+}-z_{-})g_{31}}, \nonumber \\
\mathcal{A}_{1-}&=&i\mathcal{C}_{1+}=\frac{[1+(1+\beta)z_{-}]\digamma}{(z_{+}-z_{-})g_{31}}, \nonumber \\
\mathcal{B}_{1-}&=&i\mathcal{B}_{1+}=\frac{\sqrt{2}(1-i)(1+\beta)z_{-}\digamma}{2(z_{+}-z_{-})g_{31}}, \nonumber \\
\mathcal{C}_{1-}&=&i\mathcal{A}_{1+}=\frac{[i-1-(2+\beta+i\beta)z_{-}]\digamma}{2(z_{+}-z_{-})g_{31}},
\end{eqnarray}
\\
with $z_{\pm}=[(i-1)(2+\beta)\pm i\sqrt{2(\beta^{2}+4\beta+2)}]/2$. We have checked that these analytical expressions can be used to generate the same results as shown in Fig.~\ref{figA3}.

\begin{figure*}
\begin{centering}
\includegraphics[width=0.96\textwidth]{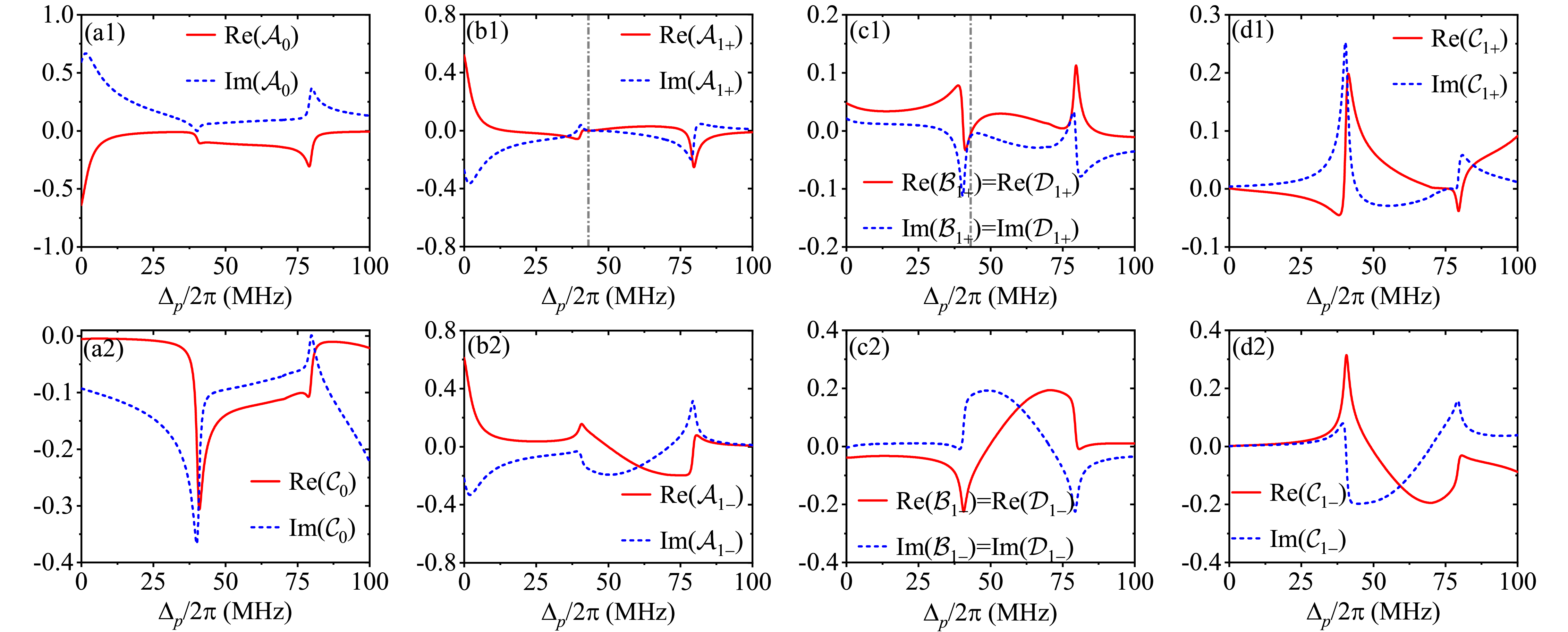}
\caption{Elements $\mathcal{A}_{0}$ (a1), $\mathcal{A}_{1+}$ (b1), $\mathcal{B}_{1+}=\mathcal{D}_{1+}$ (c1), and $\mathcal{C}_{1+}$ (d1) as well as $\mathcal{C}_{0}$ (a2), $\mathcal{A}_{1-}$ (b2), $\mathcal{B}_{1-}=\mathcal{D}_{1-}$ (c2), and $\mathcal{C}_{1-}$ (d2) in the coefficient matrix $\hat{X}$ vs probe detuning $\Delta_{p}$ attained with the same parameters as in Fig.~\ref{fig5}.}\label{figA4}
\end{centering}
\end{figure*}

Finally, choosing $\Delta_{c}=-\Delta_{d}=60$ MHz and $\phi=\pi/4$ as in Fig.~\ref{fig5}, we plot relevant elements of the coefficient matrix $\hat{X}$ against probe detuning $\Delta_{p}$ in Fig.~\ref{figA4}. It is not difficult to find that $\mathcal{A}_{0}=\mathcal{C}_{0}$, $\mathcal{A}_{1-}=i\mathcal{C}_{1+}$, $\mathcal{B}_{1-}=i\mathcal{B}_{1+}$, and $\mathcal{C}_{1-}=i\mathcal{A}_{1+}$ don't hold any more so that we further have $|\tilde{r}_{ps}^{+-}|\ne|\tilde{r}_{ps}^{-+}|$ in addition to $|\tilde{r}_{pp}^{+-}|\ne|\tilde{r}_{pp}^{-+}|$ and $|\tilde{t}_{ps}^{+-}|\ne|\tilde{t}_{ps}^{-+}|$. Above discussions based on a truncated transfer matrix $\hat{M}$ indicate that nonreciprocal scattering behaviors in our double-$\Lambda$ atomic system can be attributed to the non-Hermitian nonlinear interactions between a probe and a signal fields. To be more specific, it is the non-Hermitian interplay of linear $\{\mathcal{A}_{0},\mathcal{A}_{1\pm},\mathcal{C}_{0},\mathcal{C}_{1\pm}\}$ and nonlinear $\{\mathcal{B}_{0},\mathcal{B}_{1\pm},\mathcal{D}_{0},\mathcal{D}_{1\pm}\}$ contributions in respective reflection and transmission amplitudes that leads to the intriguing nonreciprocal results in Fig.~\ref{fig3}--Fig.~\ref{fig5} (see also Eqs.~(\ref{equ11}) and (\ref{equ12}) with related discussions).

\end{document}